\documentclass[useAMS,usenatbib]{mn2e}
\usepackage{graphicx}
\usepackage{rotating,times,pictex,graphicx,latexsym}
\usepackage{color}
\usepackage{longtable,amsmath}
\usepackage{lscape}
\usepackage{array}
\usepackage{lipsum}
\newcolumntype{C}[1]{>{\centering\arraybackslash}p{1.1cm}}

\title[Analysis of 13 Clusters]{Properties of Star Clusters -- III: Analysis of 13 FSR Clusters using UKIDSS-GPS and VISTA-VVV}

\author[Buckner \& Froebrich]{
Anne S.M Buckner,$^{1}$\thanks{E-mail: asmb2@kent.ac.uk}
Dirk Froebrich,$^{1}$\thanks{E-mail: df@star.kent.ac.uk}
\\
$^{1}$ Centre for Astrophysics and Planetary Science, University of Kent, Canterbury, CT2 7NH, United Kingdom\\
}

\date{Accepted XXX. Received YYY; in original form ZZZ}

\pubyear{2016}

\begin{document}
\label{firstpage}
\pagerange{\pageref{firstpage}--\pageref{lastpage}}
\maketitle

\begin{abstract}

Discerning the nature of open cluster candidates is essential for both
individual and statistical analyses of cluster properties. Here we establish the
nature of thirteen cluster candidates from the FSR cluster list using photometry
from the 2MASS and deeper, higher resolution UKIDSS-GPS and VISTA-VVV surveys.
These clusters were selected because they were flagged in our previous studies
as expected to contain a large proportion of pre-main sequence members or are at
unusually small/large Galactocentric distances. We employ a decontamination
procedure of $JHK$ photometry to identify cluster members. Cluster properties
are homogeneously determined and we conduct a cross comparative study of our
results with the literature (where available). Seven of the here studied
clusters were confirmed to contain PMS stars, one of which is a newly confirmed
cluster. Our study of FSR1716 is the deepest to date and is in notable
disagreement with previous studies, finding that it has a distance of about
7.3\,kpc and age of 10\,--\,12\,Gyr. As such, we argue that this cluster is a
potential globular cluster candidate.

\end{abstract}




\section{Introduction}\label{sect_intro}

Star clusters are the building blocks of the Galaxy and are tracers of both
stellar and Galactic evolution. Individually they act as laboratories,
demonstrating how stellar systems comprised of various masses work and interact
as member stars share similar properties (distance, age, reddening and
metallicity). Collectively clusters provide insight into the chemical and
structural evolution of the Galaxy.

When a large sample of clusters is available, objects of interest such as
massive clusters, old clusters near the Galactic Centre and/or young clusters
containing a large proportion of Pre-Main Sequence (PMS) members become
available for detailed study. Over the last decade the rate of cluster discovery
has significantly increased. This can be attributed to the advent of multiple
large scale Near-Infrared (NIR) and Mid-Infrared (MIR) surveys, such as GLIMPSE
\citep{2003PASP..115..953B}, 2MASS \citep{2006AJ....131.1163S}, UKIDSS-GPS
\citep{2008MNRAS.391..136L}, VISTA-VVV \citep{2010NewA...15..433M}, WISE
\citep{2010AJ....140.1868W}. Resultantly, large photometric-only cluster
candidate catalogues have become readily available e.g.
\citet{2005ApJ...635..560M}, \citet{2007MNRAS.374..399F},
\citet{2011A&A...532A.131B}, \citet{2015NewA...34...84C}. 

Unfortunately, difficulties lie in ascertaining the nature and fundamental
properties of these candidate clusters in the absence of spectroscopic and/or
astrometric data. Additional difficulties are found in the heterogeneous nature
of the catalogues themselves which are often compiled from the literature  (e.g.
WEBDA\footnote{https://www.univie.ac.at/webda/},
or DAML02 \citep{2002A&A...389..871D}). It has been argued that incongruity
between individual cluster properties is not problematic, so long as the
reliability of the values and/or methods used to derived them is considered
(e.g. \citet{2006MNRAS.371.1641P}, \citet{2015arXiv150801296N}). However, this
is impractical for the compilation of a large cluster catalogues, and ultimately
any global analyses undertaken therewith would need to be treated with great
care due to the resulting heterogeneity of cluster property values. To ensure
the validity of large scale analyses undertaken with cluster samples, it is
therefore essential that their properties are homogeneously derived, so that any
uncertainties in the determined values are systematic. Consistency of properties
derived by different authors is a primary concern, particularly for clusters
that are sparsely populated and less well defined on the field (i.e. lack
prominent features such as a strong Main Sequence (MS), giants etc.). Attempts
have been made by the community to statistically address this issue by
developing methodologies that homogeneously derive the properties of cluster
samples, but as these rely on modelled cluster sequences and/or positional data,
the accuracy of individual cluster property values remains questionable
(\citet{2010A&A...516A...2M}, \citet{2012A&A...543A.156K},
\citet{2015A&A...576A...6P}).

Our aim in this series of papers has been to homogeneously investigate the
fundamental properties and large scale distribution of Galactic open clusters.
In  \citet{2013MNRAS.436.1465B} (Paper I, hereafter) we established a foreground
star counting technique as a distance measurement and presented an automatic
calibration and optimisation method for use on large samples of clusters with
NIR photometric data only. We combined this method with colour excess
calculations to determine distances and extinctions of objects in the FSR list
cluster sample from \citet{2007MNRAS.374..399F} and investigated the $H$-band
extinction per kpc distance in the Galactic Disk as a function of Galactic
longitude. In total, we determined distance estimates to 771, and extinctions
values for 775, open cluster candidates from the FSR list.

In \citet{2014MNRAS.444..290B} (Paper II, hereafter) we investigated the
relationship between scale height and cluster age. We homogeneously derived
cluster ages and developed a novel approach to calculate cluster scale heights,
which significantly reduced established constraints on sample size. Applying our
scale height method to the homogeneous MWSC catalogue by
\citet{2013A&A...558A..53K}, the DAML02 list by \citet{2002A&A...389..871D} and
the WEBDA database, we were able to trace the scale height evolution of clusters
in detail for the first time, finding a marked increase in scale height at 1
Gyr. We also determined the parameters of 298 open cluster candidates, of which
we confirmed 82 as real clusters for the first time. 

Following that analysis, it became apparent that some objects in the FSR list
warranted further investigation as they were either suspected to contain a large
number of PMS members, or are old clusters near the Galactic Centre (GC). In all
cases, these clusters have either not been previously analysed in the literature
or only with low resolution 2MASS photometry. For both types of clusters the
problem is essentially the same: their Colour-Colour Magnitude (CCM) diagrams
lack the necessary detectable age defining features to accurately fit
theoretical cluster isochrones as only their brightest member stars are visible
above the 2MASS $K$-band detection limit, an effect which is exaggerated at
small Galactic longitudes as stellar crowding becomes more prominent in the
survey. Obviously, the properties derived/refined for these clusters in the
literature are therefore questionable, as with variable distance and extinction
values multiple isochrones can appear to be a good fit. For example, the
age-defining red giant clump of open cluster candidate FSR1716 (see
Sect.\,\ref{cluster_1716}) is below the 2MASS $K$-band detection limit, and
resultantly its properties are strongly disputed in the literature. Despite
\citet{2010MNRAS.409.1281F} and \citet{2008A&A...491..767B} both fitting Solar
metallicity isochrones to the CCM diagrams of FSR1716, there is no agreement in
its properties with a distance estimate of $0.8$\,-\,$7.0$\,kpc, age estimate of
$2$\,-\,$7$\,Gyr, or to whether the cluster is, in fact, a globular cluster.
Hence, an analysis of these clusters using deeper, higher resolution photometry
is needed to accurately derive their properties and to confirm their true
nature. In this paper we present that analysis utilising photometry from the
UKIDSS-GPS and VISTA-VVV surveys.

This paper is structured as follows. In section \,\ref{sect_methods} we present
our cluster sample and analysis methods. The results for each individual cluster
including a comparative analysis of previously derived properties (where
available), is presented in section \,\ref{sect_results}. We discuss and
conclude our findings in section \,\ref{sect_conclude}.

\section{Analysis Methods} \label{sect_methods}

In this section we present our cluster sample, and our analysis methods which
are based on photometric archival data and isochrone fitting.

\subsection{Cluster Selection}\label{sect_select}

Following our analyses of the FSR list clusters in Papers I\,$\&$\,II with 2MASS
photometry, it became apparent that some objects warranted further
investigation. These clusters were either (i) suspected to contain a large
proportion of PMS members or (ii) have unusually small/large Galactocentric
distances and/or distances from the Sun with respect to the remainder of the FSR
List. For the former, deeper magnitude photometry is necessary to confirm that
these clusters' members are predominantly PMS, and to derive accurate properties
for the clusters as only their brightest member stars were visible above the
2MASS $K$-band detection limit.  

For the latter, clusters were found to either be within $5$\,kpc of, or around
$13$\,kpc away from, the GC\footnote{Assuming a Solar Galactocentric distance of
$R^{\odot}_{GC}=8.00$\,kpc \citep{2012arXiv1202.6128M}}, or have a distance from
the Sun in excess of $10$\,kpc (the FSR List peaks at a distance of about
$3$\,kpc). Inclusion of deeper, higher resolution photometry will therefore make
visible the dimmer cluster members which are on the MS but below the 2MASS
detection limit, thus enabling more accurate isochrone fits to be made and
ultimately the clusters' distances can be confirmed or revised.

Based on our Paper II study, 19 clusters in the FSR list sample were flagged as
potentially having a large proportion of PMS members, and 5 clusters were
flagged as having notable Galactocentric distances. Of these photometry from the
UKIDSS-GPS and VISTA-VVV surveys was available for 13 of the clusters: FSR0089,
FSR0188, FSR0195, FSR0207, FSR0301, FSR0636, FSR0718, FSR0794, FSR0828, FSR0870,
FSR0904, FSR1189 and FSR1716.

\subsection{Photometry and Cluster Radii}\label{sect_radii}

Clusters' central coordinates are taken from \citet{2007MNRAS.374..399F}. The
core radius of each cluster, $r_{cor}$, is determined from a radial star density
profile fit of the form:

\begin{equation}\label{eq_density1}
\rho(r)=\rho_{bg}+\rho_{cen} \left[ 1 + \left(\frac{r}{r_{cor}}\right)^2 
\right]^{-1}
\end{equation}

Where $r$ is the distance from the cluster centre; $\rho(r)$ the projected
radial star density; $\rho_{bg}$ is the projected background star density which
is assumed constant; and $\rho_{cen}$ the central star density above the
background of the cluster.

JHK photometry was extracted from the UKIDSS-GPS, VISTA-VVV and 2MASS NIR point
source catalogues in a circular $0.3^\circ$ area around the clusters' central
coordinates. The apparent core radii for our selected clusters is
$<0.01^{\circ}$, thus an area of $0.3^{\circ}$ satisfactorily encompasses the
members of each cluster. 

To optimise the quantity and quality of photometry used in this study, we make
the following selections. For the deep, high resolution VISTA-VVV and UKIDSS-GPS
surveys we select all objects that have been given a mergedClass classification
of ``-1" (i.e. object has a probability of $\ge\,90\%$ of being a star;
\citet{2008MNRAS.391..136L}), and that have photometry available in each of the
$JHK$ filters, down to their limiting $K$-band magnitudes of 18.1\,mag and
17.8\,mag respectively. As we will use the 2MASS photometry for bright sources
only (see Sect.\,\ref{sect_ccm_construct}) we select the highest photometric
quality with a Qflag of ``AAA", (i.e. detected by the $JHK$ filters with a
S\,:\,N of greater than 10 and corrected photometric uncertainties of less than
0.11\,mag; \citet{2006AJ....131.1163S}) down to a $K$-band magnitude of
12.0\,mag.

\subsection{Cluster Membership Identification}\label{sect_mpi}

Ideally cluster members should be automatically identified through their colours
and magnitudes. This should be done without the need to fit  isochrones to
Colour Magnitude or Colour-Colour diagrams or relying on manual, subjective
selection of cluster members. This series of papers uses an approach that is
based on the works of \citet{2007MNRAS.377.1301B} and
\citet{2010MNRAS.409.1281F}, known as the Photometric Decontamination Technique
(PDT). The PDT is the approach of choice as it determines the likelihood that a
star $i$ projected onto a cluster is a member based on its position in CCM
space.  

To distinguish true members from interloping field stars we define the cluster
area, $A_{cl}$, as a circular area around the centre of a cluster within which
the majority of members are expected to be contained. The radius of this cluster
area is two times the cluster core radius ($2\,\times\,r_{cor}$). We also define
a control area, $A_{con}$, as a ring around the centre of a cluster within which
all stars are expected to belong to the field population. The inner radius of
this ring is five times the cluster core radius ($5\,\times\,r_{cor}$) and the
outer radius is $0.3^{\circ}$. For a discussion of the validity of our cluster
and control area selections the reader is referred to the relevant sections in
Papers\, I and II. 

To determine the likelihood, $P^i_{cl}$, that star $i$ is a cluster member we
start by estimating the CCM distance, $r^{i,j}_{CCM,cl}$, between star $i$ and
every other star $j$ ($i \ne j $) in the cluster area using the following
equation:

\begin{equation}\label{eq_rccm_cl}
r^{i,j}_{CCM,cl}=\sqrt{\frac{1}{2}\left( J_i - J_j \right)^2 + \left( JK_i - JK_j
\right)^2 + \left( JH_i - JH_j \right)^2} 
\end{equation}

where $i$ and $j$ identify all stars in the cluster area, $J$ refers to the
$J$-band magnitudes of the stars, and $JK$ and $JH$ to the respective colours of
them. This combination of colours and magnitudes has been shown to be the most
effective in separating cluster members from field stars over a wide range of
cluster ages (\citet{2004A&A...415..571B}, \citet{2007MNRAS.377.1301B}).

We then identify the CCM distance $r^{i,N}_{CCM,cl}$ for star $i$ that
corresponds to the $N^{th}$ nearest neighbour in CCM space in the cluster area.
For our analysis here we set $N=15$. As shown in Paper\,I, the specific choice
value of $N$ is essentially arbitrary and does not significantly affect the
identification of the most likely cluster members. It essentially defines the
balance between the `resolution' in CCM space at which potential cluster members
can be separated from field stars and the S:N of the determined membership
likelihood. In other words increasing $N$ will lead to a higher S/N for
$P^i_{cl}$ but cluster features (such as the MS) traced by the most likely
cluster members in CCM space will be less well defined. 

We then determine the CCM distance $r^{i,j}_{CCM,con}$ between star $i$ in the
cluster area and every other star $j$ in the control area adapting
Eq.\,\ref{eq_rccm_cl} where now $j$ identifies all stars in the control area. We
then count the number of stars $N^{i,con}_{CCM}$ in the control area that have
$r^{i,j}_{CCM,con} < r^{i,N}_{CCM,cl}$.  From this we can estimate the
likelihood that star $i$ is a member of the cluster by:

\begin{equation}\label{eq_pcl}
P^i_{cl}=1.0-\frac{N^{i,con}_{CCM}}{N}\frac{A_{cl}}{A_{con}}.
\end{equation}

where $P^i_{cl}$ is the \textit{Membership Probability Index} (MPI) of star $i$;
and $A_{cl}$ and $A_{con}$ are the areas covered by the cluster and control
fields, respectively. In principle, $P^i_{cl}$ can have a negative value due to
statistical fluctuations of the number of field stars in the control and cluster
area. Thus it is in fact not a true probability and hence has been named a
probability index instead. As a negative value simply means that a star is very
unlikely to be a member of the cluster, all negative $P^i_{cl}$ values are set
to zero.  

In principle stellar members can be statistically identified from (and
photometric MPIs augmented by) positional data as it is reasonable to expect
there to be less contamination from interlopers towards a cluster's centre, thus
the likelihood that a star $i$ is a member of the cluster increases with
decreasing radial distance (\citet{2012A&A...539A.125D},
\citet{2014A&A...561A..57K}). However, membership probabilities derived in this
way should be treated with caution as they can be unreliable for clusters that
are: (i) dense, as stellar crowding in their central regions will make accurate
membership determination difficult; (ii) projected onto high density stellar
field, as not clearly distinguished from the field, and a significant proportion
of true members may be outside the determined cluster radius; (iii) young, as
these clusters may not necessarily be circular in projection and have
substructure (e.g. \citet{2009ApJ...696.2086S}, \citet{2015MNRAS.448.2504G}). As
the clusters studied in this work each fall into one or more of these
categories, we do not use spatial probabilities to identify members. For a full
discussion the reader is referred to Paper\,II and \citet{2010MNRAS.409.1281F}.

\subsection{Construction of CCM Diagrams} \label{sect_ccm_construct}

Cluster $J-K$ vs. $K$ Colour-Magnitude Diagrams (CMD) and $H-K$ vs $J-H$
Colour-Colour Diagrams (CCD) were constructed using photometry from the
UKIDSS-GPS/VISTA-VVV surveys for point sources at $K\,<\,12$\,mag, and 2MASS for
point sources brighter than $K\ge12$\,mag. To ensure no point sources at
$K\sim$12\,mag which appear in both UKIDSS-GPS/VISTA-VVV and 2MASS were plotted
twice, we cross-referenced the surveys and removed point sources accordingly.
Stellar memberships for each cluster were determined as described in
Sect.\,\ref{sect_mpi} and overplotted on the diagrams. We define the extinction
law for each cluster from the reddening free parameter Q for each star in the
cluster area, where

\begin{equation}\label{eq_qparam}
\begin{array}{lcl}
Q=JH-\left(\chi\cdot HK\right) \text{\,\,\,\,\,\,\,\,and\,\,\,\,\,\,\,\,} \chi = \frac{E(J-H)}{E(H-K)}\\
\end{array}
\end{equation}

As described in Paper I, a value of $Q\,\le\,-0.05$\,mag by more than
$1\,\sigma$ (estimated from the photometric uncertainties) denotes a Young
Stellar Object.

It has been shown in the literature the value of $\chi$ typically varies between
1.55 \citep{1990ARA&A..28...37M} and 2.00 \citep{2008BaltA..17..253S}. Using the
methods described above and in Sect.\,\ref{sect_meth_iso}, we identify each
clusters most likely members and make manual isochrone fits assuming a value of
$\chi=1.55$ until a satisfactory fit is made. In cases where our assumed value
of $\chi$ is clearly too low (or high) we vary the conversion factors, $C_{KH}$
and $C_{JH}$, until the best fit for the cluster is found. Where 

\begin{equation}\label{eq_x_cjhk}
\begin{array}{rcl}
\frac{A_{K}}{A_{H}} = C_{KH} \text{\,\,\,\, and \,\,\,\,\,} \frac{A_{J}}{A_{H}} = C_{JH} \text{\,\,\,\,\,\,\,\,therefore\,\,\,\,\,\,\,\,} \chi = \frac{C_{JH}-1}{1-C_{KH}} \\
\end{array}
\end{equation}

and $A_J$, $A_H$, $A_K$ are the NIR $J$-, $H$- and $K$-band extinctions
respectively.

\subsection{Isochrone Fitting} \label{sect_meth_iso}

The clusters studied in this paper lack spectroscopic measurements, without
which there is no way of determining their metallicities. In Paper II we showed
that the median metallicity of clusters in the WEBDA catalogue is $Z=\,0.02$
(i.e. Solar), hence it is reasonable to assume a Solar metallicity for open
clusters in the Solar Neighbourhood in cases where their metallicity is unknown.
It should be noted that systematic uncertainties caused by using a (slightly)
incorrect metallicity are small. For example, if a Solar metallicity is
incorrectly assumed for an open cluster of $-0.4<\,[Fe/H]\,<0.2$ there will be
an intrinsic uncertainty comparable to $log(age/yr)\sim\,0.1$.

We fit Solar metallicity Geneva \citep{2001A&A...366..538L} or PMS
\citep{2000A&A...358..593S} isochrones to the highest probability cluster
members on the CMDs and CCDs, utilising our homogeneously determined distance,
extinction and age values from Paper II as a starting point for the fits.

\section{Results}\label{sect_results}

In this section we present and discuss in detail the findings of our analysis of
each cluster using the methods outlined in Section \ref{sect_methods}. In
addition, we conduct a cross comparative study of our results with the
literature (where available). A summary of the properties and nature of each
cluster can be found in Table \ref{obsinterest_results}.

\begin{figure*}
        \includegraphics[width=\columnwidth,viewport=54 360 558 720]{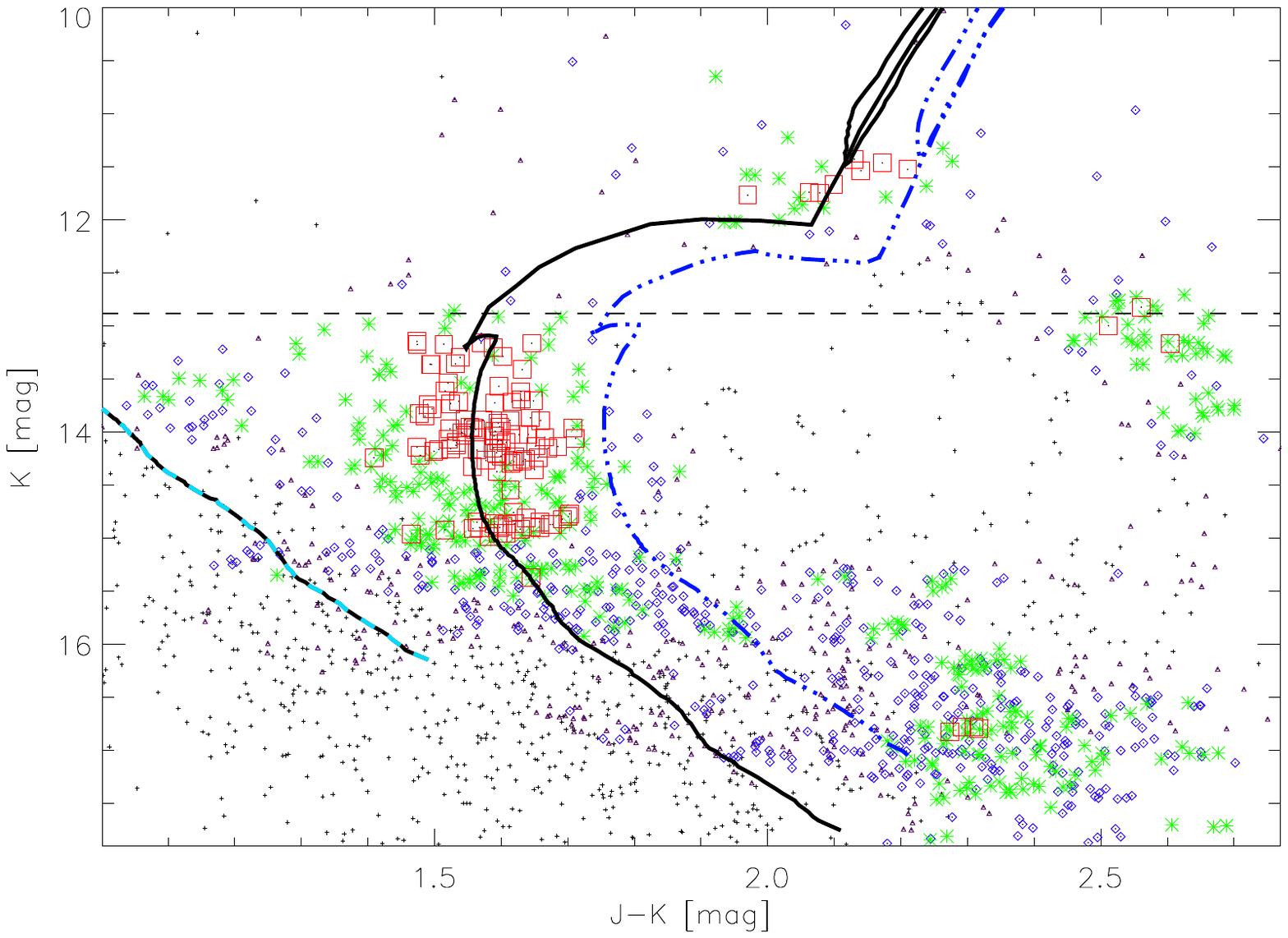} \hfill
        \includegraphics[width=\columnwidth,viewport=54 360 558 720]{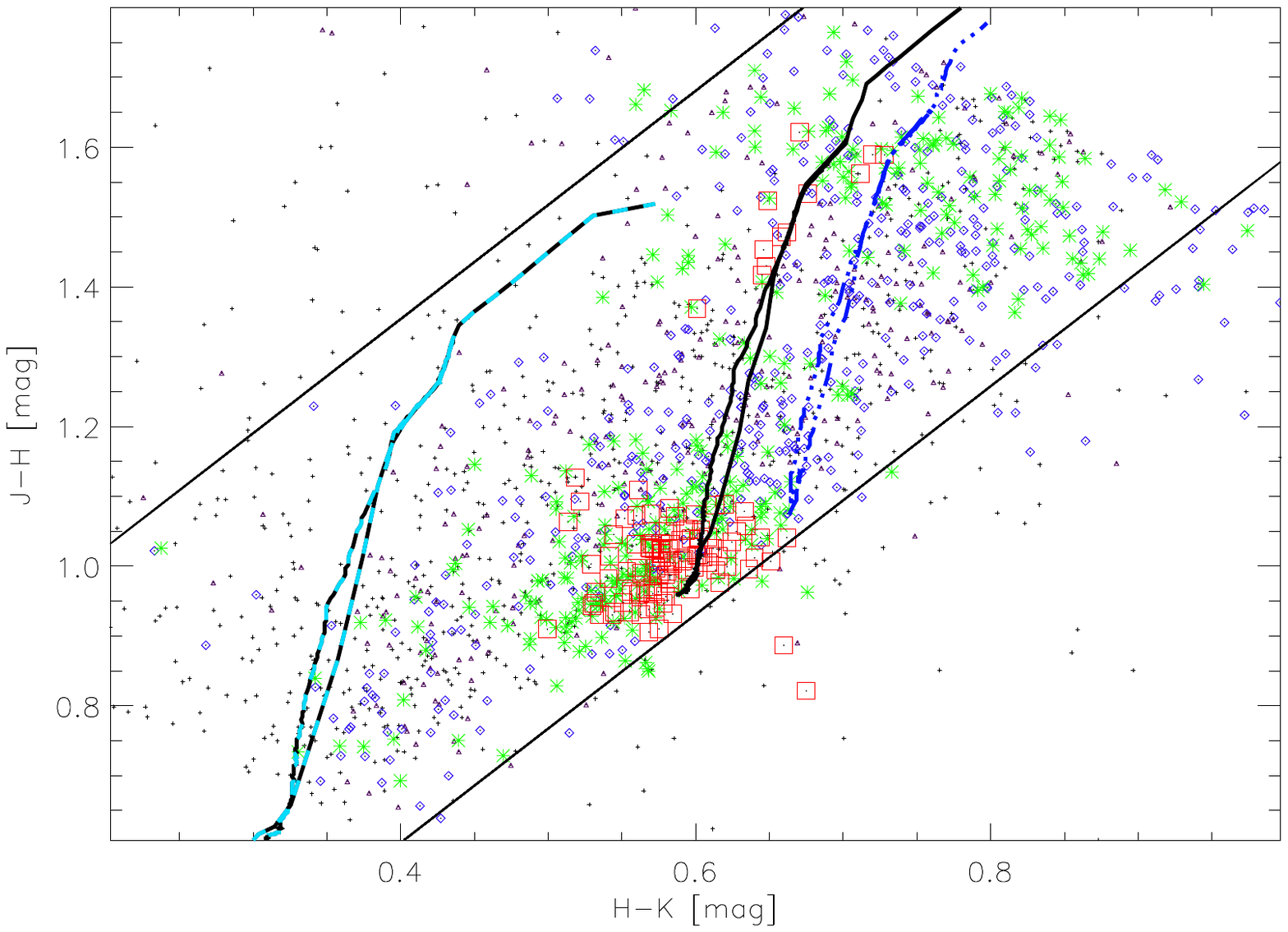}\\
        \caption{\label{fig_ccm_fsr0089} ($Left$) Colour-Magnitude and ($Right$) Colour-Colour diagrams of \textbf{FSR0089}. Photometric membership probabilities are determined for stars within $2\,\times\,r_{cor}$, with $N=15$ and are represented as follows: $P^i_{cl} > 80\%$ red squares; $60 \leq P^i_{cl} < 80\% $ green stars; $40 \leq P^i_{cl} < 60\% $ blue diamonds; $20\leq P^i_{cl} < 40\% $ purple triangles; $P^i_{cl} < 20\% $ black plus signs. The parallel black lines in the Colour-Colour plot represent the reddening band of the cluster. The horizontal dashed black line marks the 2MASS $K$-band completeness limit at the cluster's coordinates. The solid black and dashed turquoise/black isochrones represent the best fit, as determined in this study and the MWSC catalogue respectively. The triple-dot-dash blue isochrone represents the best fit as determined by \citet{2008MNRAS.390.1598F}.}
\end{figure*}
\subsection{FSR0089} \label{cluster_0089}

A confirmed cluster candidate of the FSR list, located in the first Galactic
Quadrant, flagged as the nearest to the GC in Paper II with a Galactocentric
distance of $R_{GC}=3.9$\,kpc. Figure \ref{fig_ccm_fsr0089} shows an
intermediate-old cluster with a well defined MS, turn-off and giants. The
cluster's properties are redetermined as $d=\,3.10$\,kpc, age of 500\,Myr and
$A_{H}=\,1.53$\,mag. After testing different variations we found the extinction
law that best fits the cluster is $\chi=1.64$. 

The revised age and extinction values are in approximate agreement with those
derived in Paper II, but the cluster's distance estimate has halved. As such,
FSR0089 is no longer the nearest cluster to the GC in the sample, but still has
a notable Galactocentric distance of $R_{GC}=5.5$\,kpc. This discrepancy in the
distance estimate occurred because the majority of the cluster's MS is below the
2MASS detection limit and thus was not detected when the cluster's fundamental
properties were initially derived. 

The revised values are in generally good agreement with the literature, albeit
with a slightly larger distance. The reader should note that the extinction
value given by the MWSC catalogue is not in agreement with the revised values or
the literature. Comparing their best-fit isochrone to the CMD the apparent
discrepancy is caused by their misidentification of field stars as cluster
members.

\begin{figure*}
        \includegraphics[width=\columnwidth,viewport=54 360 558 720]{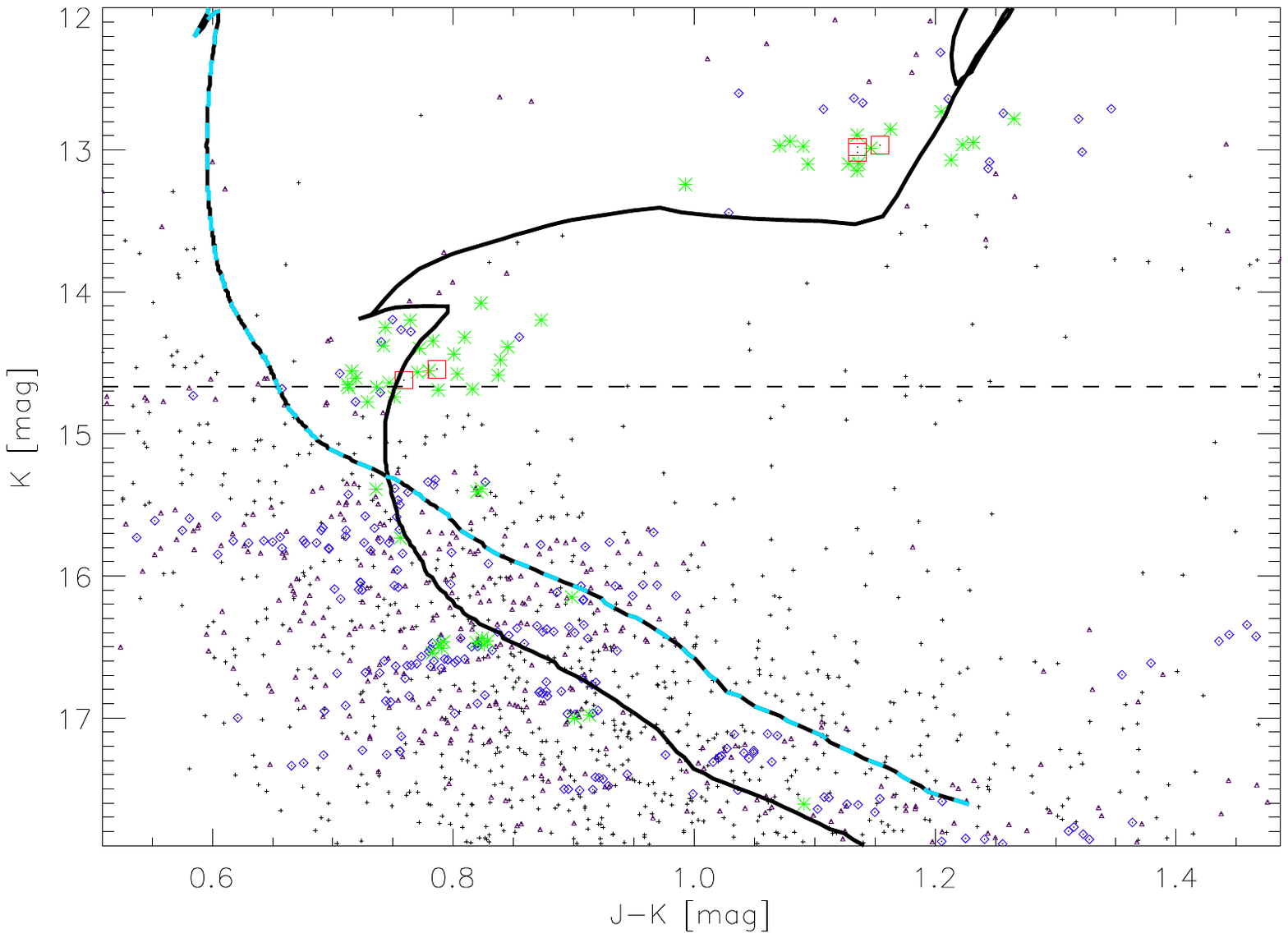} \hfill
        \includegraphics[width=\columnwidth,viewport=54 360 558 720]{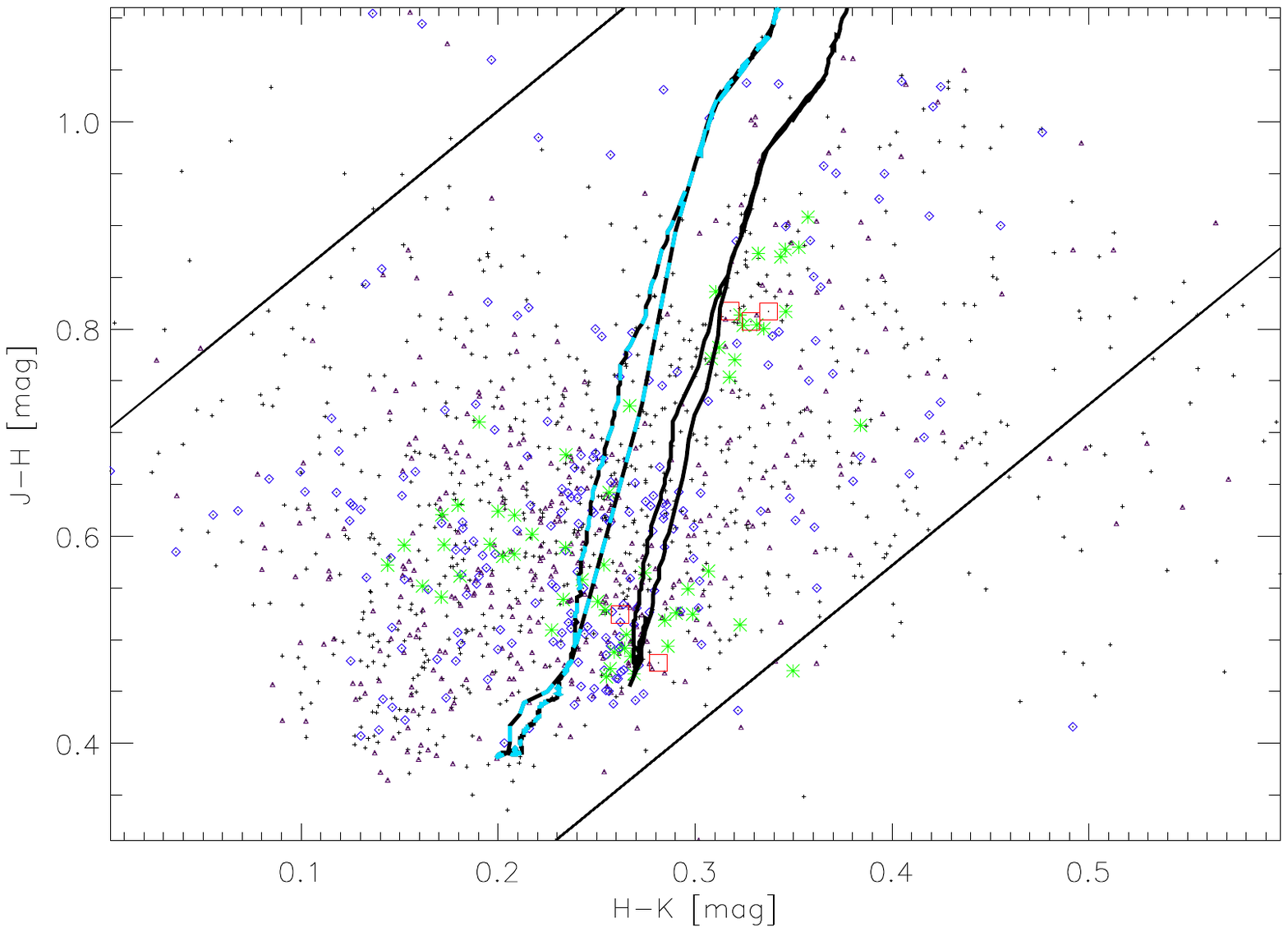}\\
        \caption{\label{fig_ccm_fsr0188} As in Figure\,\ref{fig_ccm_fsr0089}, but for \textbf{FSR0188}.}
\end{figure*}
\subsection{FSR0188}\label{cluster_0188}

A confirmed cluster candidate of the FSR list, located in the first Galactic
Quadrant. In Paper II FSR0188 was flagged as one of clusters with the largest
Solar distance with $d=\,10.50$\,kpc. Figure \ref{fig_ccm_fsr0188} shows an old
cluster with a well defined MS, turn-off and giants. The cluster has a Solar
distance of $d=\,4.90$\,kpc, age of 1\,Gyr and extinction of
$A_{H}=\,0.61$\,mag. After testing different variations we found the extinction
law that best fits the cluster is $\chi=1.55$. 

The revised age and extinction values are in general agreement with those
derived in Paper II, but the cluster's distance estimate has halved. As such,
FSR0188 is no longer one of the furthest clusters in the sample. The cause of
the distance discrepancy is that the majority of cluster sequence was below the
2MASS detection limit, which impacted on accuracy of the distance estimate; it
is a good example of the necessity for deeper magnitude photometry when deriving
cluster properties through isochrone fitting. 

FSR0188 has been previously studied by \citet{2013A&A...558A..53K} as part of
the MWSC catalogue using 2MASS photometry. The authors determined the cluster to
be younger, nearer and redder than the revised values suggest.
Figure\,\ref{fig_ccm_fsr0188} shows the CCM diagrams of the cluster with two
sets of isochrones over-plotted: those depicting the revised values, and those
given by \citet{2013A&A...558A..53K}. Clearly, the revised values are a better
fit for the cluster's sequence, the majority of which is below the 2MASS
detection limit. Again, the most likely cause of the discrepancy with the MWSC
catalogues values is their misidentification of field stars as cluster members
(in particular the giants).

\begin{figure*}
        \includegraphics[width=\columnwidth,viewport=54 360 558 720]{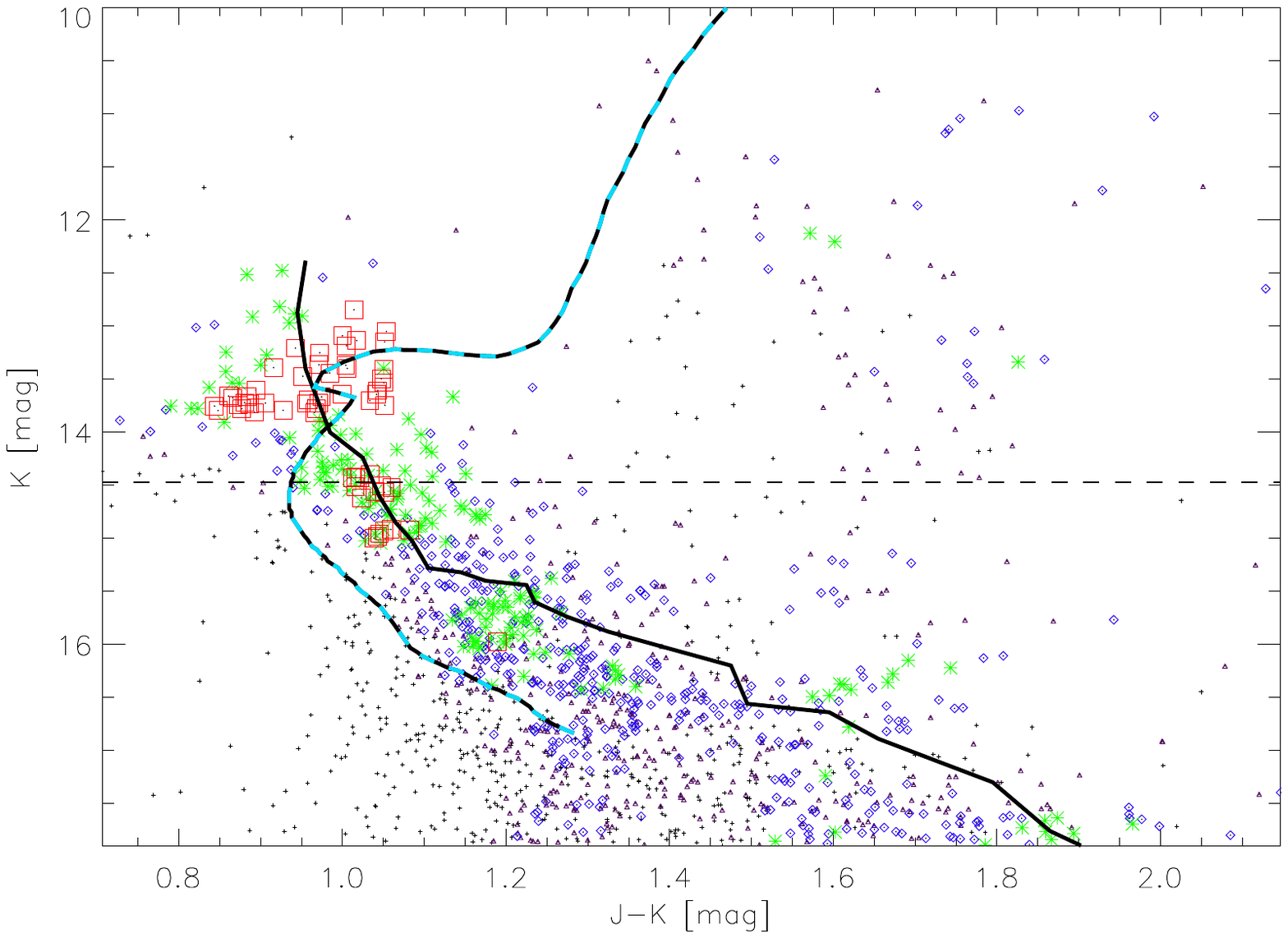}  \hfill
        \includegraphics[width=\columnwidth,viewport=54 360 558 720]{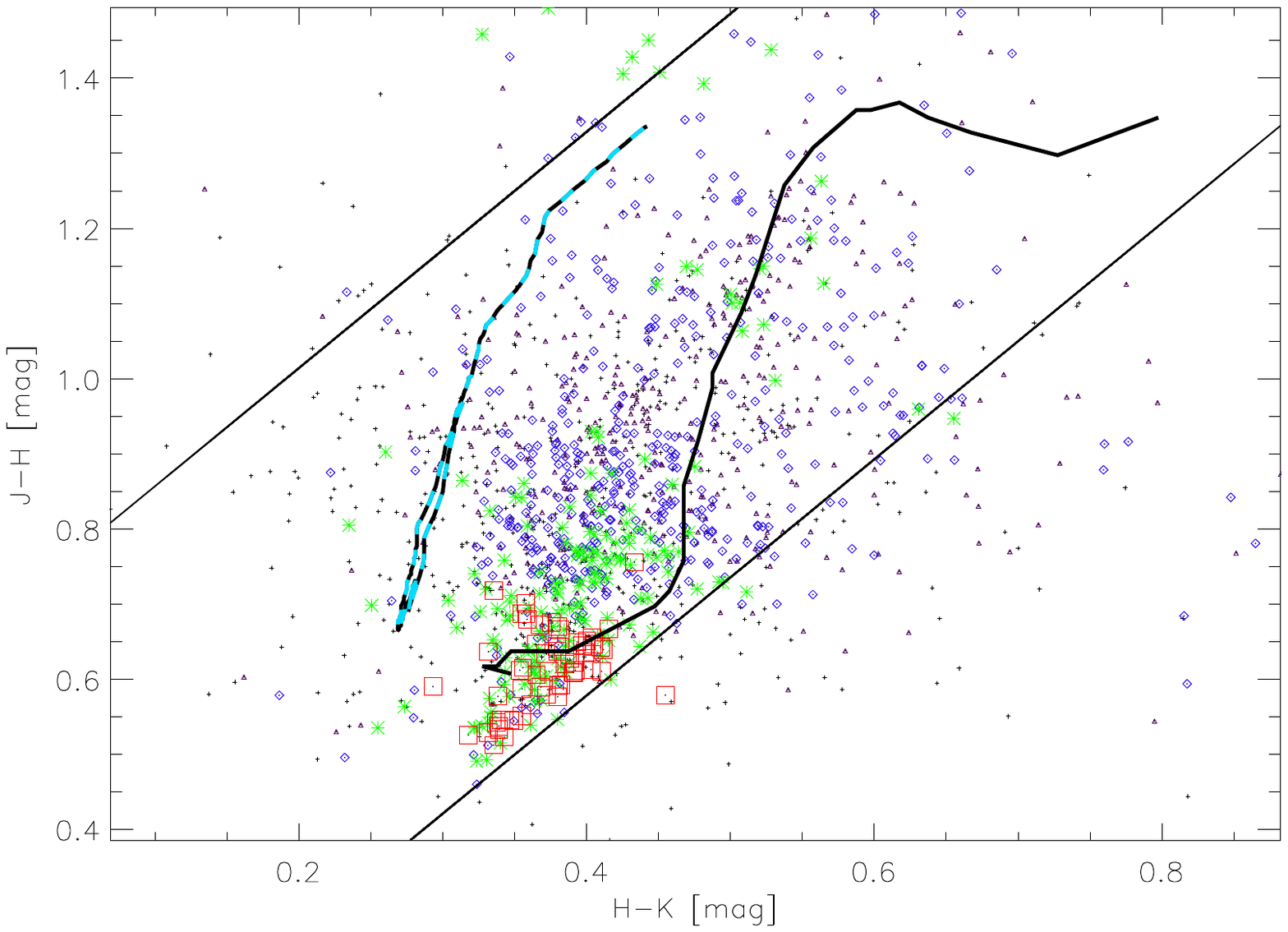} \\
        \caption{\label{fig_ccm_fsr0195} As in Figure\,\ref{fig_ccm_fsr0089}, but for \textbf{FSR0195}.}
\end{figure*}

\begin{figure*}
        \includegraphics[width=\columnwidth,viewport=54 360 558 720]{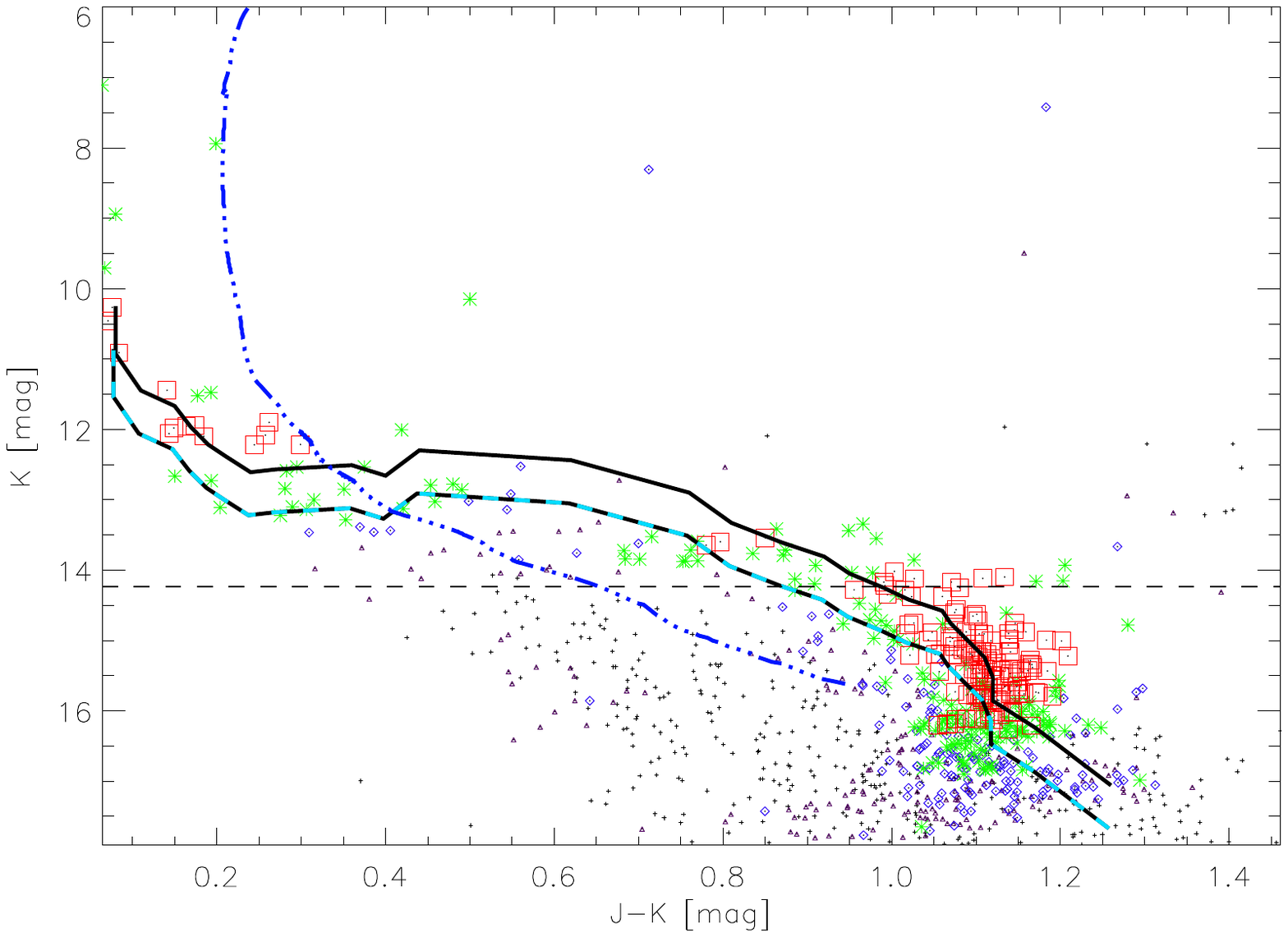}\hfill
        \includegraphics[width=\columnwidth,viewport=54 360 558 720]{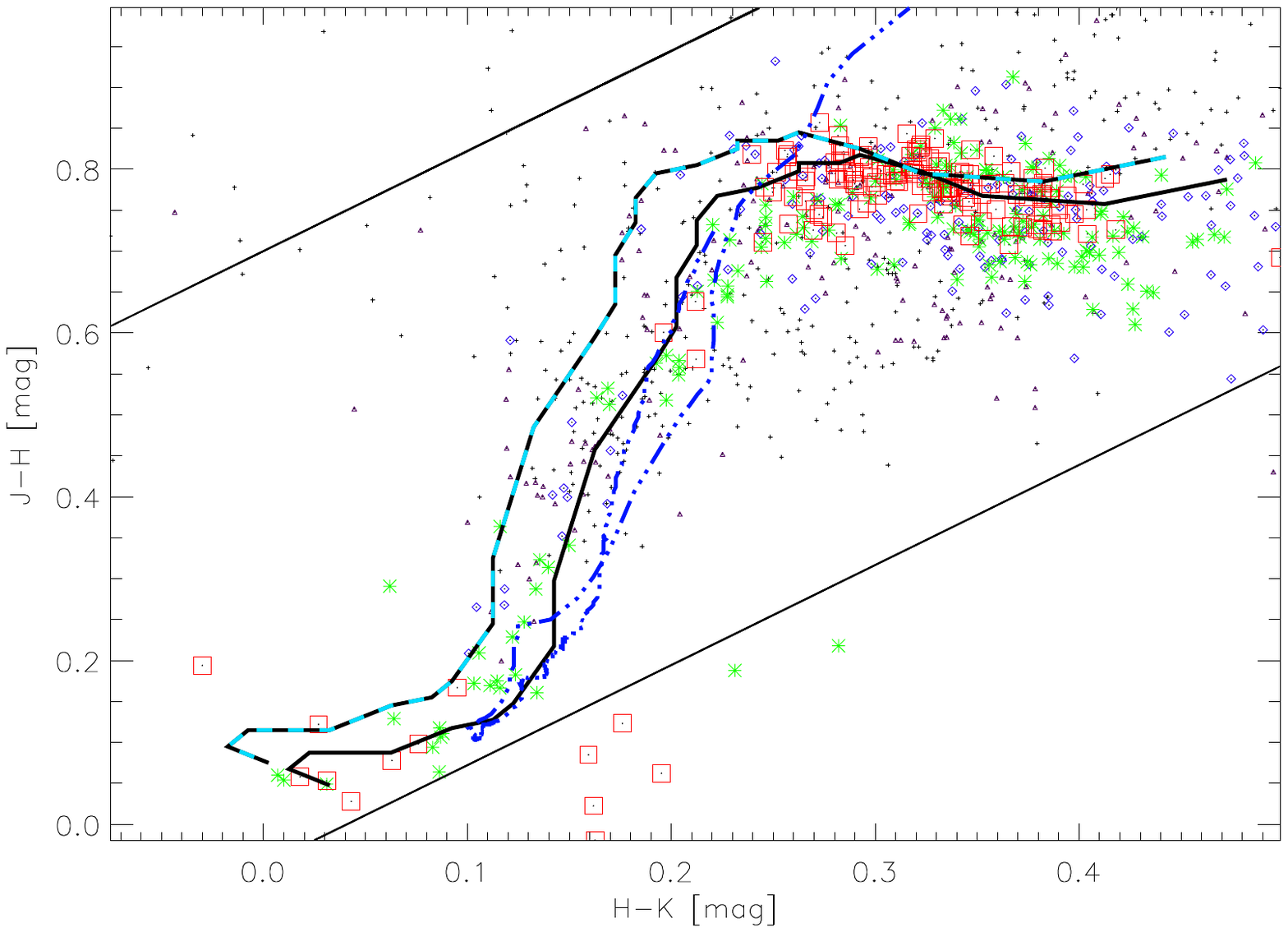}\\
        \caption{\label{fig_ccm_fsr0207} As in Figure\,\ref{fig_ccm_fsr0089}, but for \textbf{FSR0207}. The triple-dot-dash blue isochrone represents the best fit as determined by \citet{1995yCat.7092....0L}.}
\end{figure*}

\subsection{FSR0195}\label{cluster_0195}

A confirmed cluster candidate of the FSR list, located in the first Galactic
Quadrant. In Paper II FSR0195 was flagged as having a (potentially) large number
of PMS members. Figure \ref{fig_ccm_fsr0195} shows a young open cluster with a
PMS track. The cluster has a Solar distance of $d=\,3.50$\,kpc, age of 30\,Myr
and a $H$-band extinction of $A_{H}=1.25$\,mag. After testing different
variations we found the extinction law that best fits the cluster is
$\chi=1.57$. These revised age value is consistent with our value determined in
Paper II. This revised distance is approximately 1.5 times greater than that
derived in Paper II, as the addition of dimmer stars below the 2MASS detection
limit has enabled its value to be more accurately constrained. 

The parameters of FSR0195 have previously been derived by
\citet{2013A&A...558A..53K} who found the cluster to have smaller extinction and
distance values but a markedly larger age estimate of 2.2\,Gyr.
Fig.\,\ref{fig_ccm_fsr0195} clearly shows that the cluster is much younger than
this value and therefore the values given by the MWSC catalogue are inaccurate.
FSR0195 is a good example of the need for deep, high resolution photometry when
deriving clusters' fundamental properties.

\subsection{FSR0207} \label{cluster_0207}

A previously known cluster (IC\,4996), located in the first Galactic Quadrant in
the Cygnus constellation. Figure \ref{fig_ccm_fsr0207} shows a young open
cluster with a PMS track. The redetermined properties of FSR0207 are
$d=\,1.40$\,kpc, age of 10\,Myr and $A_{H}=\,0.25$\,mag. After testing different
variations we found the extinction law that best fits the cluster is
$\chi=1.22$. These revised values are in general agreement with the values
derived in Paper II.

The presence of PMS stars in FSR0207 has previously been confirmed through
photometric and spectral analysis (see e.g. \citet{1999AJ....118.1759D},
\citet{1998AJ....116.1801D}, \citet{2004IAUS..224..353Z},
\citet{2006A&A...457..237Z}, \citet{2007BASI...35..383B}). Using
SIMBAD\footnote{http://simbad.u-strasbg.fr/simbad/} we searched the cluster area
of FSR0207 for bright stars which are likely members of the cluster, finding
there is a $\beta$ Lyr type eclipsing binary (MPI\,$=\,0.59$) and a B3 type star
(MPI\,$=\,0.71$). A spectroscopic study by \citet{2011BASI...39..517M} showed
that the age of the B3 star is 8\,Myr, which is consistent with our derived
cluster age.

A search of the literature reveals that the cluster has been extensively studied
and it is generally agreed to have an age of $\sim\,10$\,Myr, Solar distance of
$\sim\,1.90$\,kpc and $H$-band extinction of $\sim\,0.35$\,mag, i.e. further and
more reddened than the revised values presented here suggest (see Table
\ref{obsinterest_results}). However, it should be noted that authors
predominantly used an extinction law of $\chi=1.60$ and UBVRI photometry to
derive these values. IRAS maps have shown there to be an IR dusty shell
surrounding the cluster which is associated with the nearby Berkeley\,87
\citep{1990AZh....67.1152L}, so the UKIDSS-GPS $JHK$ photometry utilised here
has a distinct advantage, revealing the dimmer stars in the cluster sequence
which are crucial to accurately fit modelled cluster sequence isochrones to the
CCM diagrams of FSR0207 (and derive the cluster's fundamental properties). 

\begin{figure*}
        \includegraphics[width=\columnwidth,viewport=54 360 558 720]{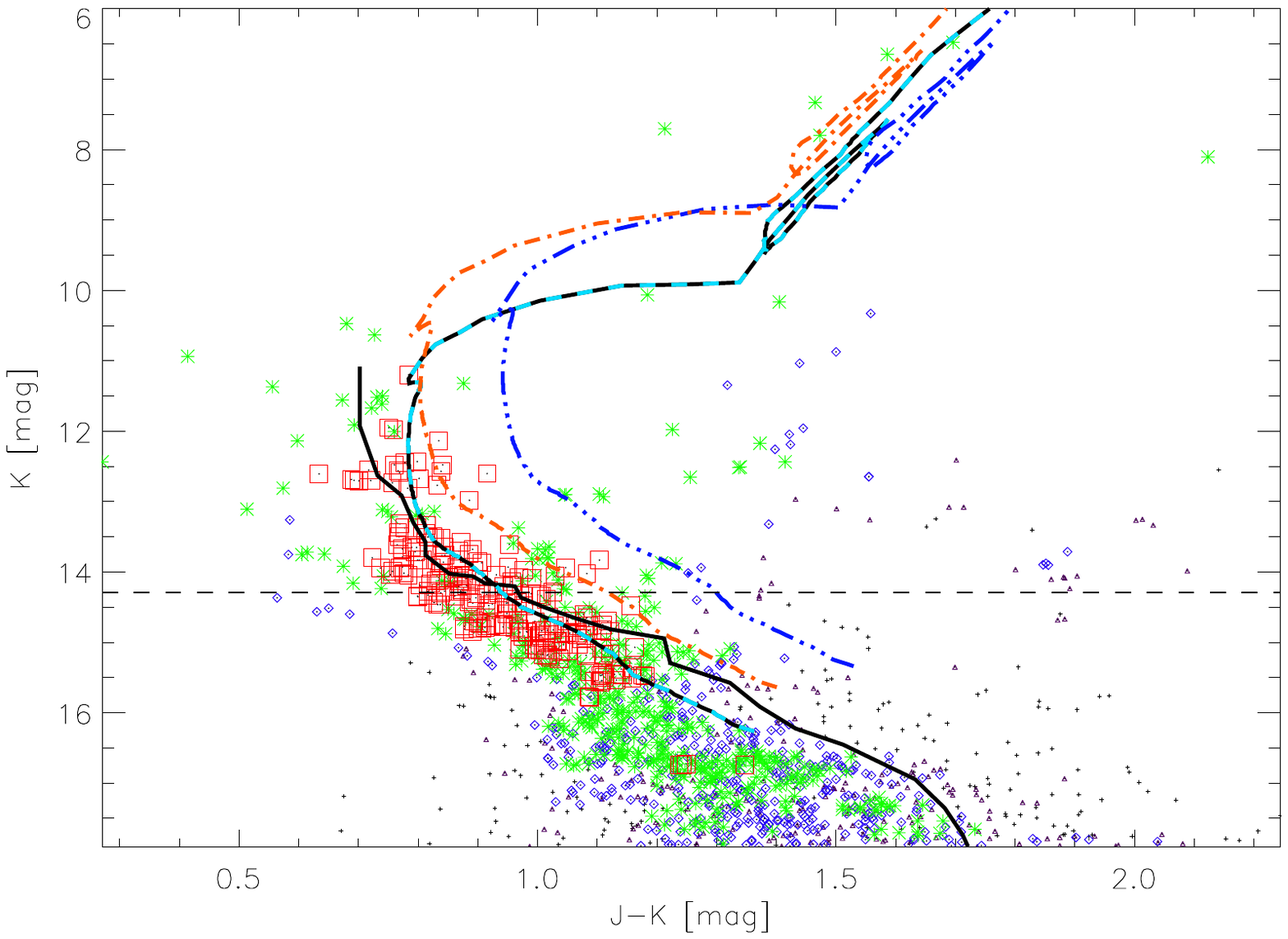}
        \includegraphics[width=\columnwidth,viewport=54 360 558 720]{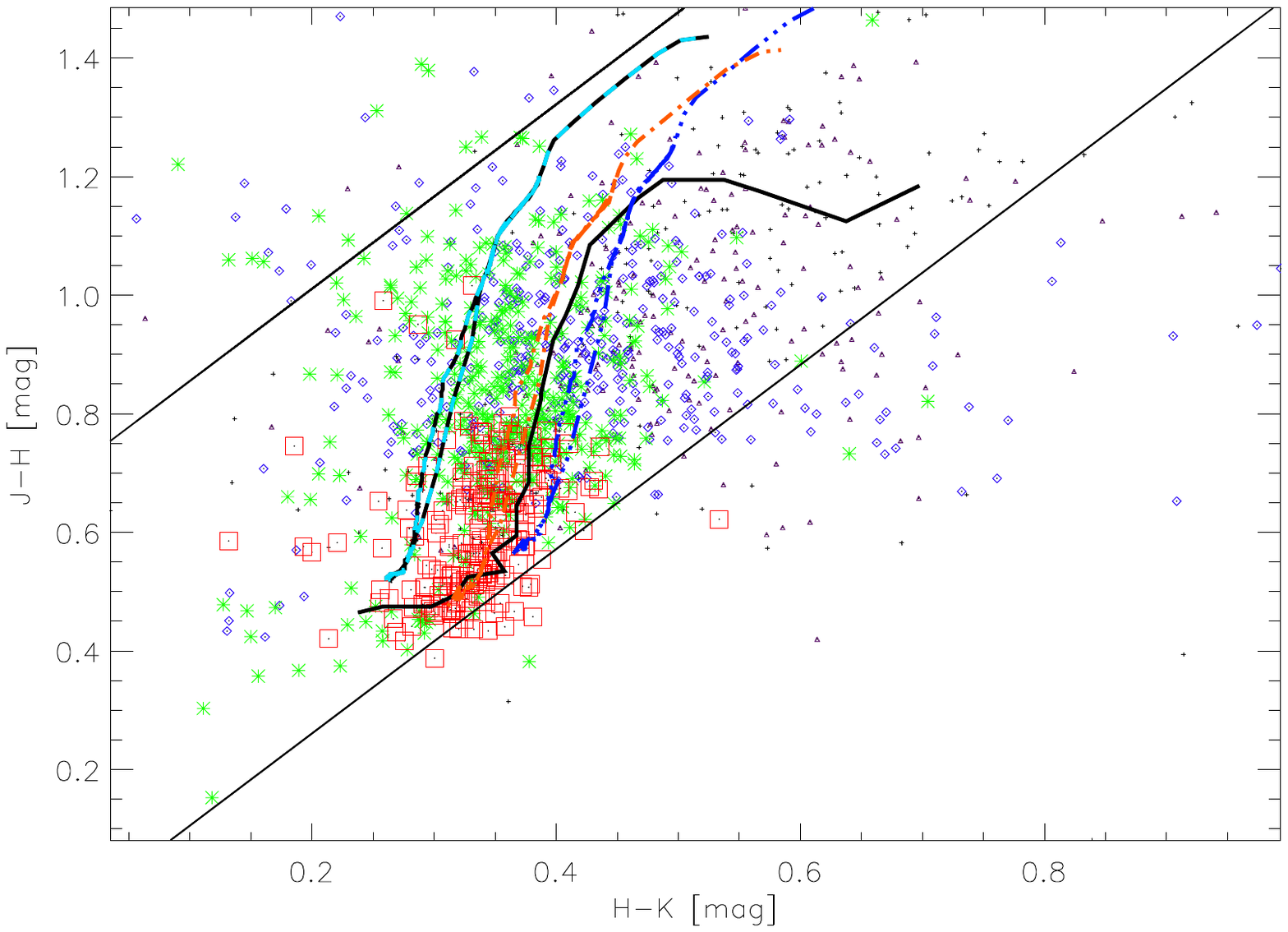}
        \caption{\label{fig_ccm_fsr0301} As in Figure\,\ref{fig_ccm_fsr0089}, but for \textbf{FSR0301}. The dot-dash orange and triple-dot-dash blue isochrones represent the best fit as determined by \citet{2008MNRAS.389..285T} and \citet{2007A&A...467.1065M} respectively.}
\end{figure*}

\begin{figure*}
        \includegraphics[width=\columnwidth,viewport=54 360 558 720]{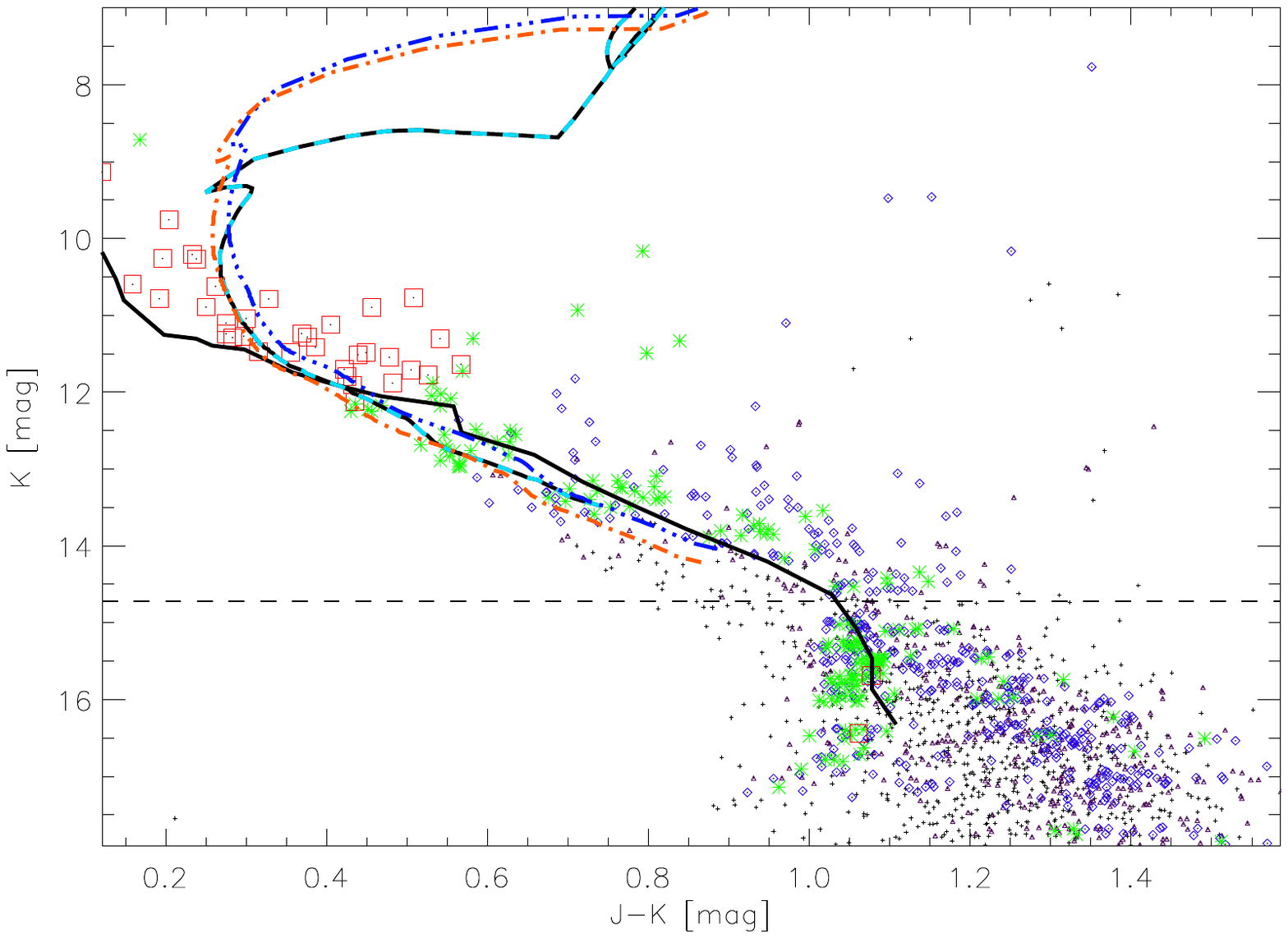} \hfill
        \includegraphics[width=\columnwidth,viewport=54 360 558 720]{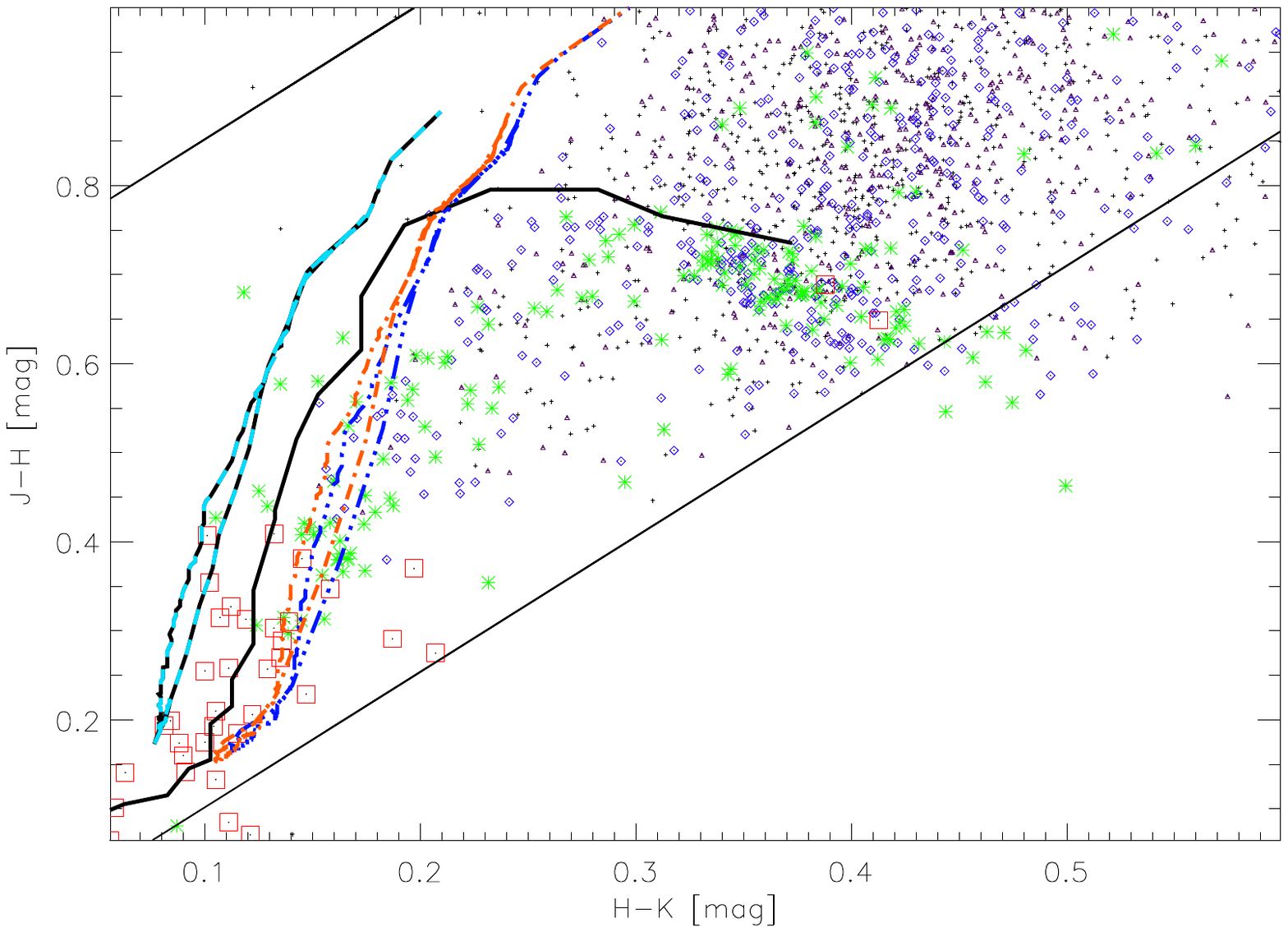} \\
        \caption{\label{fig_ccm_fsr0636} As in Figure\,\ref{fig_ccm_fsr0089}, but for \textbf{FSR0636}. The dot-dash orange and triple-dot-dash blue isochrones represent the best fit as determined by \citet{2002AJ....123..905A} and \citet{2007A&A...467.1065M}.}
\end{figure*}

\subsection{FSR0301} \label{cluster_0301}

A previously known cluster (Berkeley\,55), located in the second Galactic
Quadrant in the Cygnus constellation. Figure \ref{fig_ccm_fsr0301} shows a young
open cluster with a PMS track. The scattering of the brightest objects at the
top of the MS could be real or could be caused by  misidentification of members.
The cluster's properties are redetermined as $d=\,2.25$\,kpc, age of 50\,Myr and
$A_{H}=\,0.87$\,mag. After testing different variations we found the extinction
law that best fits the cluster is $\chi=1.55$. These revised values are in
general agreement with the values derived in Paper II, albeit giving a slightly
larger age but smaller extinction for the cluster. 

FSR0301's revised age value is in good agreement with
\citet{2012AJ....143...46N} who conducted an in depth study of the cluster using
UBVJHK photometry and $z$-band spectra. Other studies determined the cluster to
have a significantly smaller Solar distance and to be much older at
$\sim$\,300\,Myr (\citet{2008MNRAS.389..285T}, \citet{2007A&A...467.1065M} and
the MWSC catalogue), which is most likely a result of member misidentification.
However, the CCM diagrams clearly show this  300\,Myr isochrone is too old for
the cluster, which is a notably poor fit to the cluster sequence below
$K\approx13$\,mag. Furthermore, \citet{2012AJ....143...46N} showed that the
brightest  stars on the MS are B$4$\,-\,$6$, which confirms the cluster's age to
be greater than 40\,Myr but less than 100\,Myr.

\subsection{FSR0636}\label{cluster_0636}

A cluster previously known in the literature (King\,6), located in the second
Galactic Quadrant in the Local Spiral Arm. In Paper II FSR0636 was flagged as
having a (potentially)  large number of PMS members. Figure
\ref{fig_ccm_fsr0636} shows a young open cluster with a PMS track. The cluster
has a Solar distance of $d=\,0.72$\,kpc, age of 60\,Myr and extinction of
$A_{H}=\,0.33$\,mag. After testing different variations we found the extinction
law that best fits the cluster is $\chi=1.52$. These revised age and extinction
values are consistent with those previously determined in this series.

To date there have been three studies of FSR0636 in the literature and all are
in general agreement with the revised distance values but conclude the cluster
to be much older at $\ge$ \,250\,Myr  (\citet{2002AJ....123..905A},
\citet{2007A&A...467.1065M}, \citet{2013A&A...558A..53K}). As demonstrated in
Figure \ref{fig_ccm_fsr0636}, although a 250\,Myr isochrone can be fitted to the
cluster, it fails to fit the cluster sequence for objects with $K>14$\,mag
entirely, which fit a much younger 50\,Myr isochrone. Furthermore, the earliest
stars on the MS are B-stars of type B$5$\,-\,$7$ \citep{2007BaltA..16..167S},
i.e. the cluster has an age of less than 100\,Myr. Note, the MWSC derives an
extinction of $A_H =0.19$ i.e. markedly smaller than the accepted literature
value, resulting from their use of a different extinction law ($\chi=2.00$) to
that of the other literature studies ($\chi=1.60$).

\begin{figure*}
        \includegraphics[width=\columnwidth,viewport=54 360 558 720]{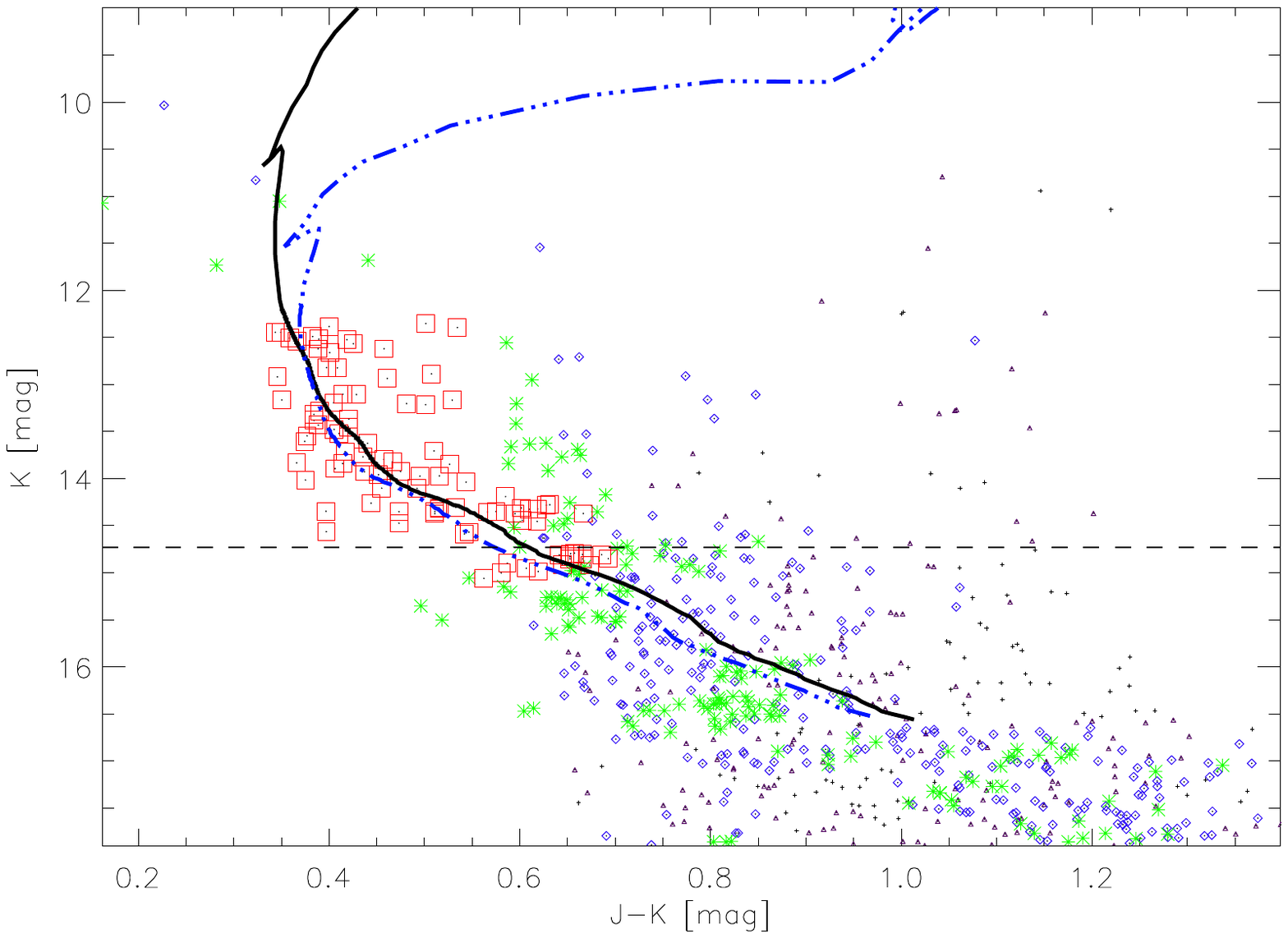}\hfill
        \includegraphics[width=\columnwidth,viewport=54 360 558 720]{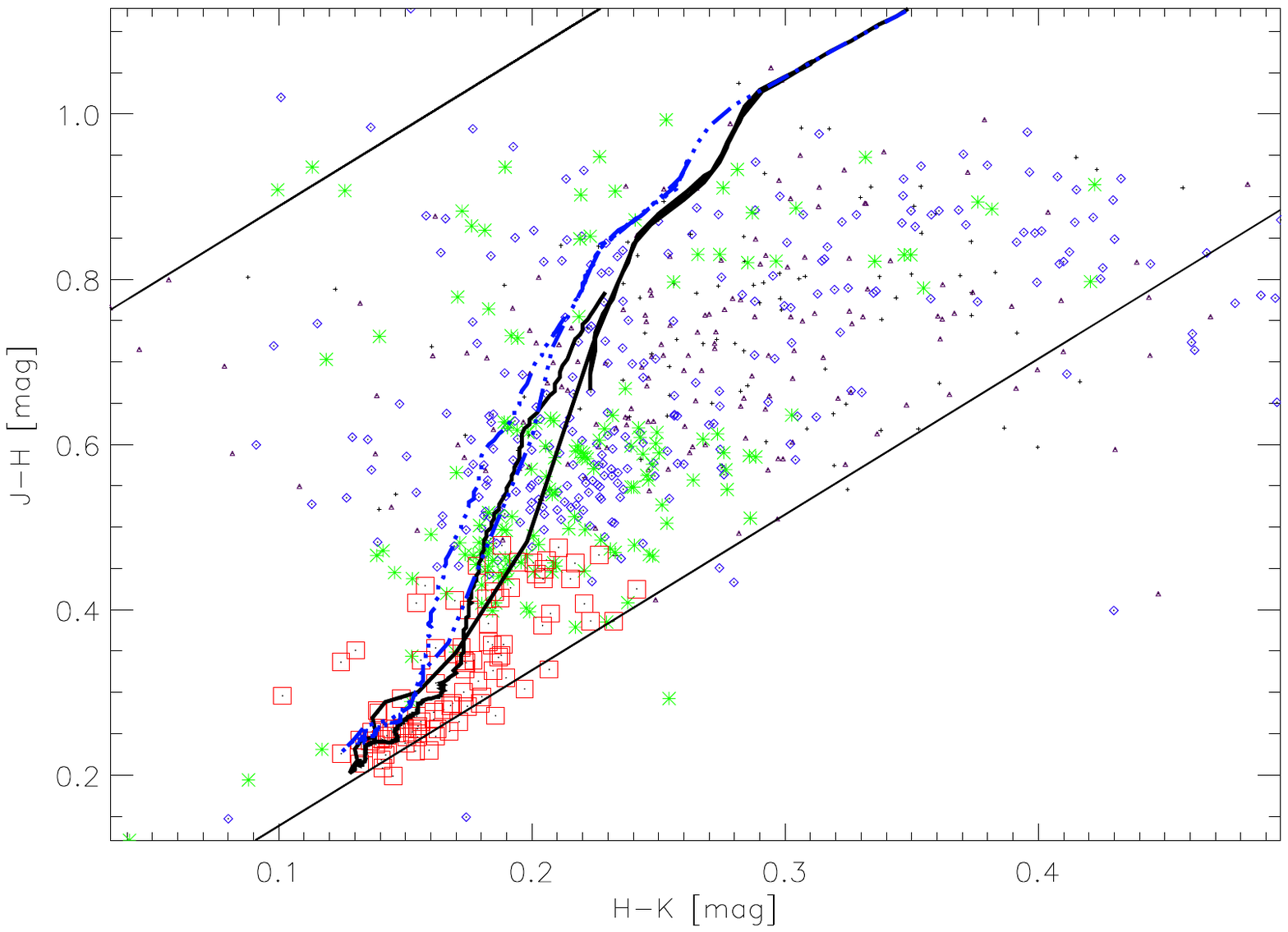}\\
        \caption{\label{fig_ccm_fsr0718} As in Figure\,\ref{fig_ccm_fsr0089}, but for \textbf{FSR0718}. The triple-dot-dash blue line represents a 300\,Myr isochrone with the same distance and extinction as determined by this study.}
\end{figure*}

\begin{figure*}
        \includegraphics[width=\columnwidth,viewport=54 360 558 720]{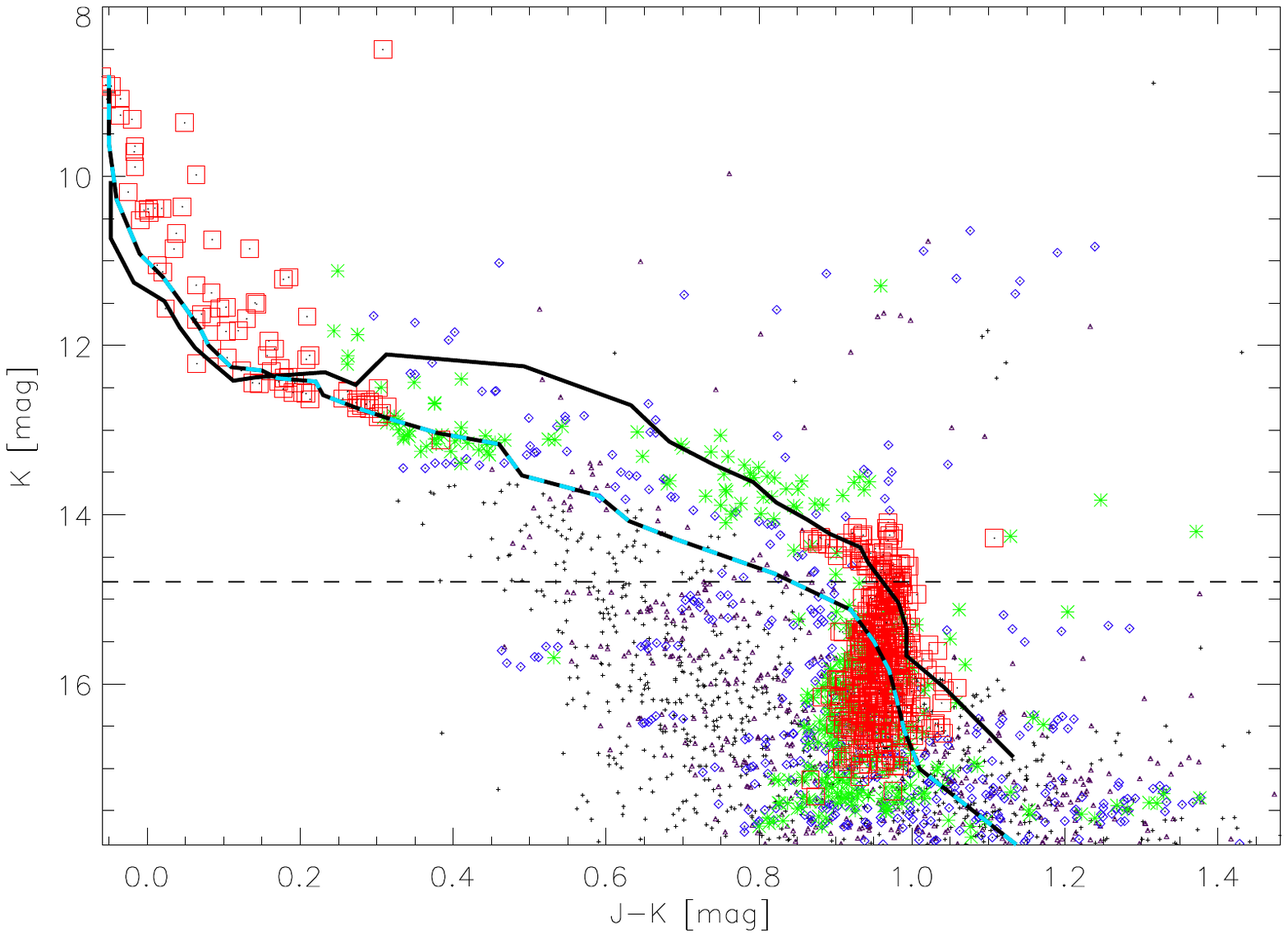}\hfill
        \includegraphics[width=\columnwidth,viewport=54 360 558 720]{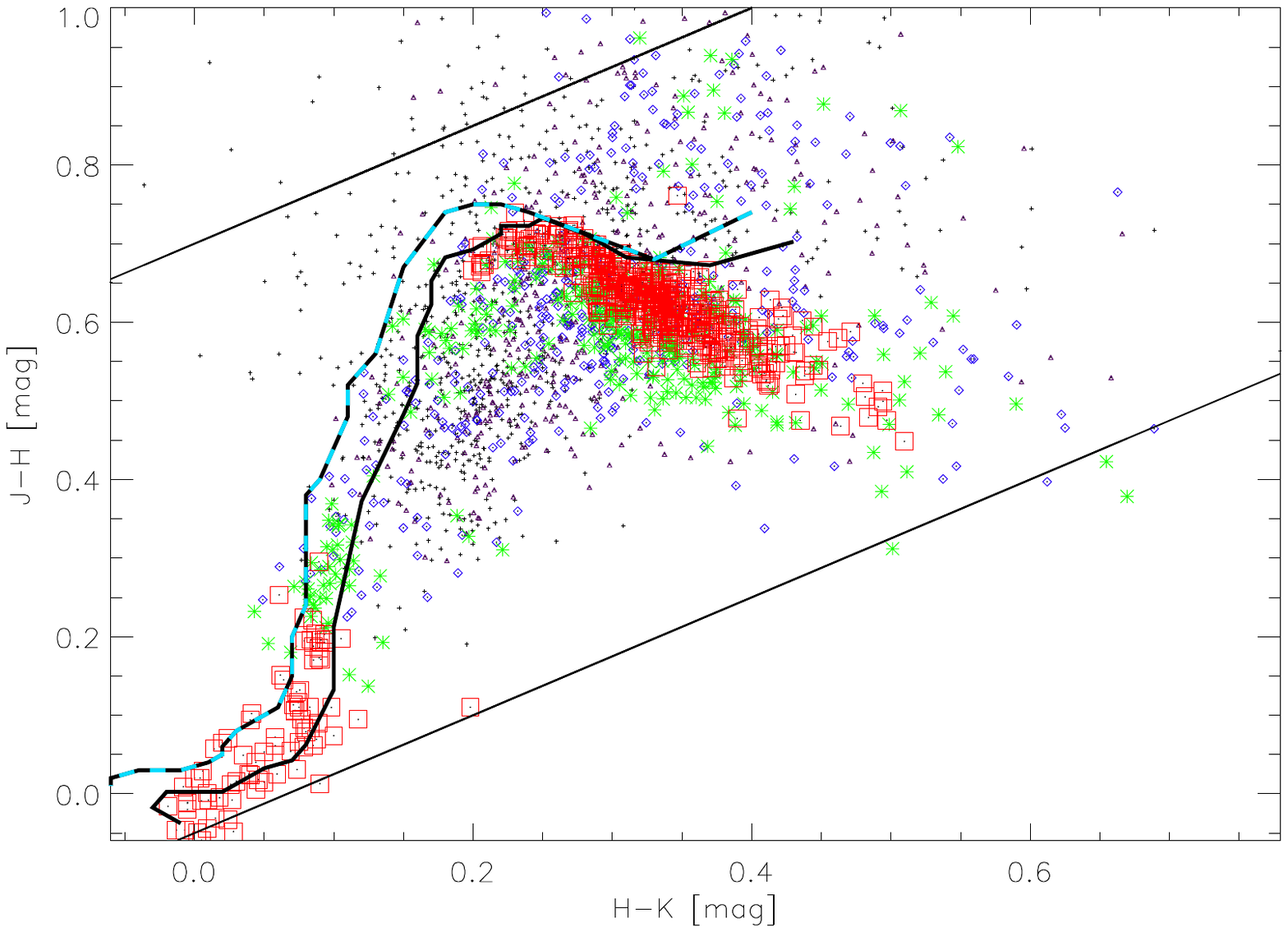}\\
        \caption{\label{fig_ccm_fsr0794} As in Figure\,\ref{fig_ccm_fsr0089}, but for \textbf{FSR0794}.}
\end{figure*}

\begin{figure*}
        \includegraphics[width=\columnwidth,viewport=54 360 558 720]{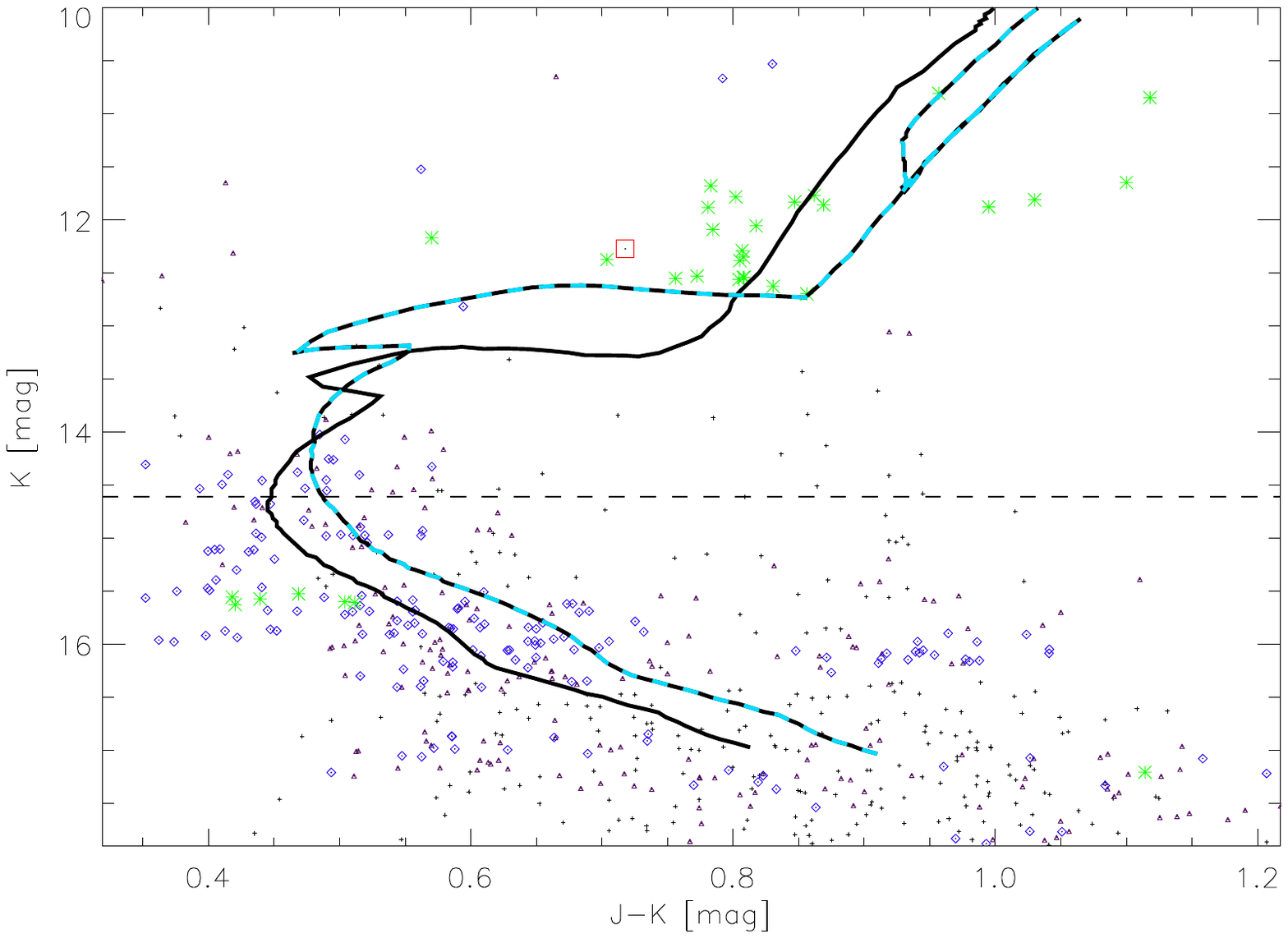}\hfill
        \includegraphics[width=\columnwidth,viewport=54 360 558 720]{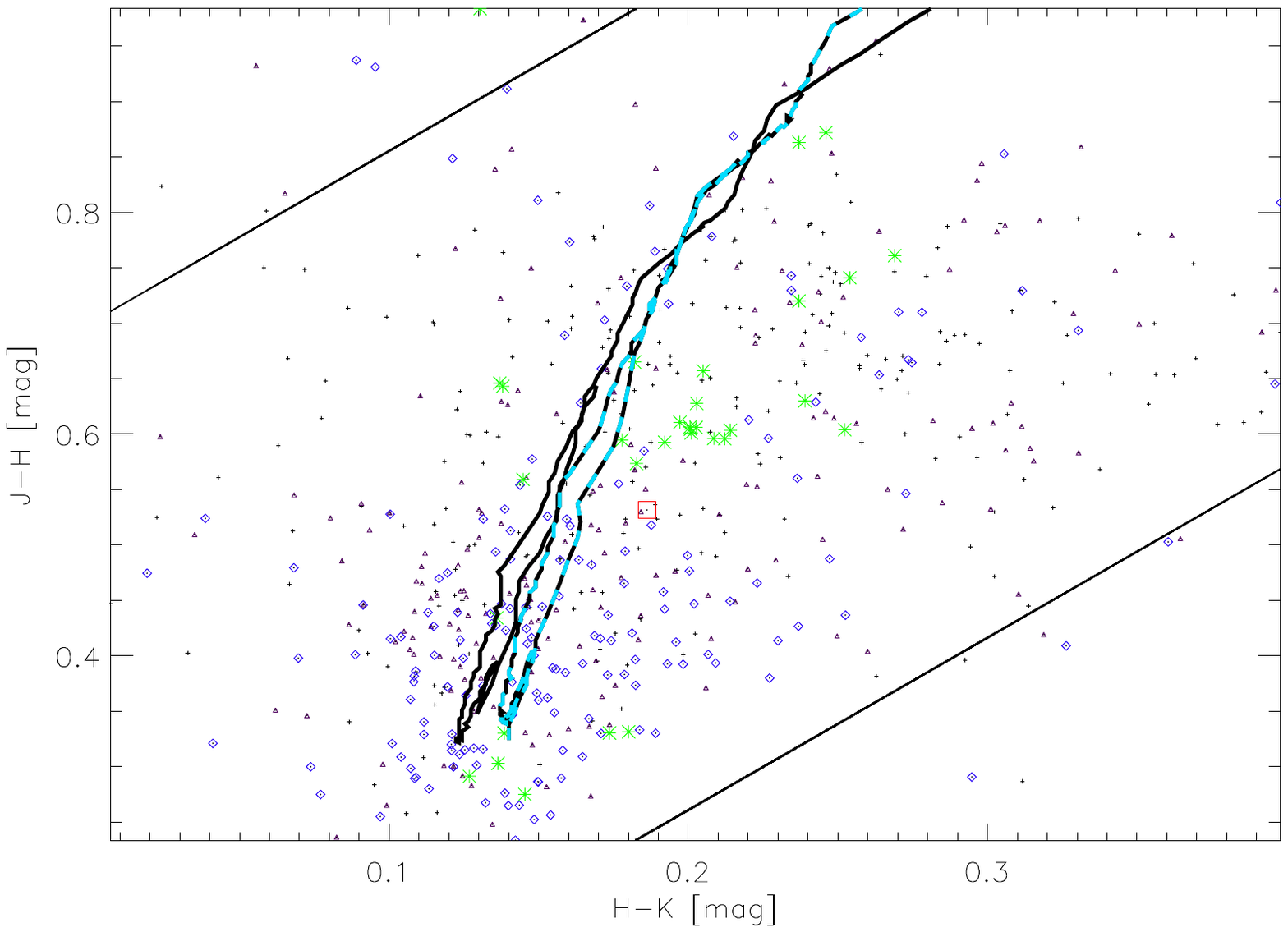}\\
        \caption{\label{fig_ccm_fsr0828} As in Figure\,\ref{fig_ccm_fsr0089}, but for \textbf{FSR0828}.}
\end{figure*}

\subsection{FSR0718}\label{cluster_0718}

A cluster previously known in the literature (Berkeley\,15), located in the
second Galactic Quadrant in the Auriga constellation. In Paper II FSR0718 was
flagged as having a (potentially)  large number of PMS members. Figure
\ref{fig_ccm_fsr0718} shows a young open cluster but without a PMS track it was
assumed to have. The cluster has a Solar distance of $d=\,2.40$\,kpc, age of
80\,Myr and extinction of $A_{H}=\,0.45$\,mag. After testing different
variations we found the extinction law that best fits the cluster is
$\chi=1.88$. These revised age and extinction values are consistent with those
previously determined in this series, albeit with a slightly smaller extinction
and larger age.

The properties of FSR0718 are strongly disputed in the literature, but most
authors derive an age in excess of $>$\,300\,Myr. This is much older than the
revised value, which is most likely a result of misidentification of field stars
(outside of $2\,\times\,r_{cor}$) as cluster members. To demonstrate, the reader
is referred to the cluster's CCM diagrams in Figure\,\ref{fig_ccm_fsr0718},
where it is shown that if there were giant members at $K\le12$\,mag and
$0.9$\,mag$\le\,[J-H]\le\,1.3$\,mag, the cluster would have an age in excess of
$\sim$\,300\,Myr. As observed by \citet{2010MNRAS.401..621T} further analysis in
needed to discern the earliest member spectral types in FSR0718 for absolute
clarification of the cluster's age.

\begin{figure*}
        \includegraphics[width=\columnwidth,viewport=54 360 558 720]{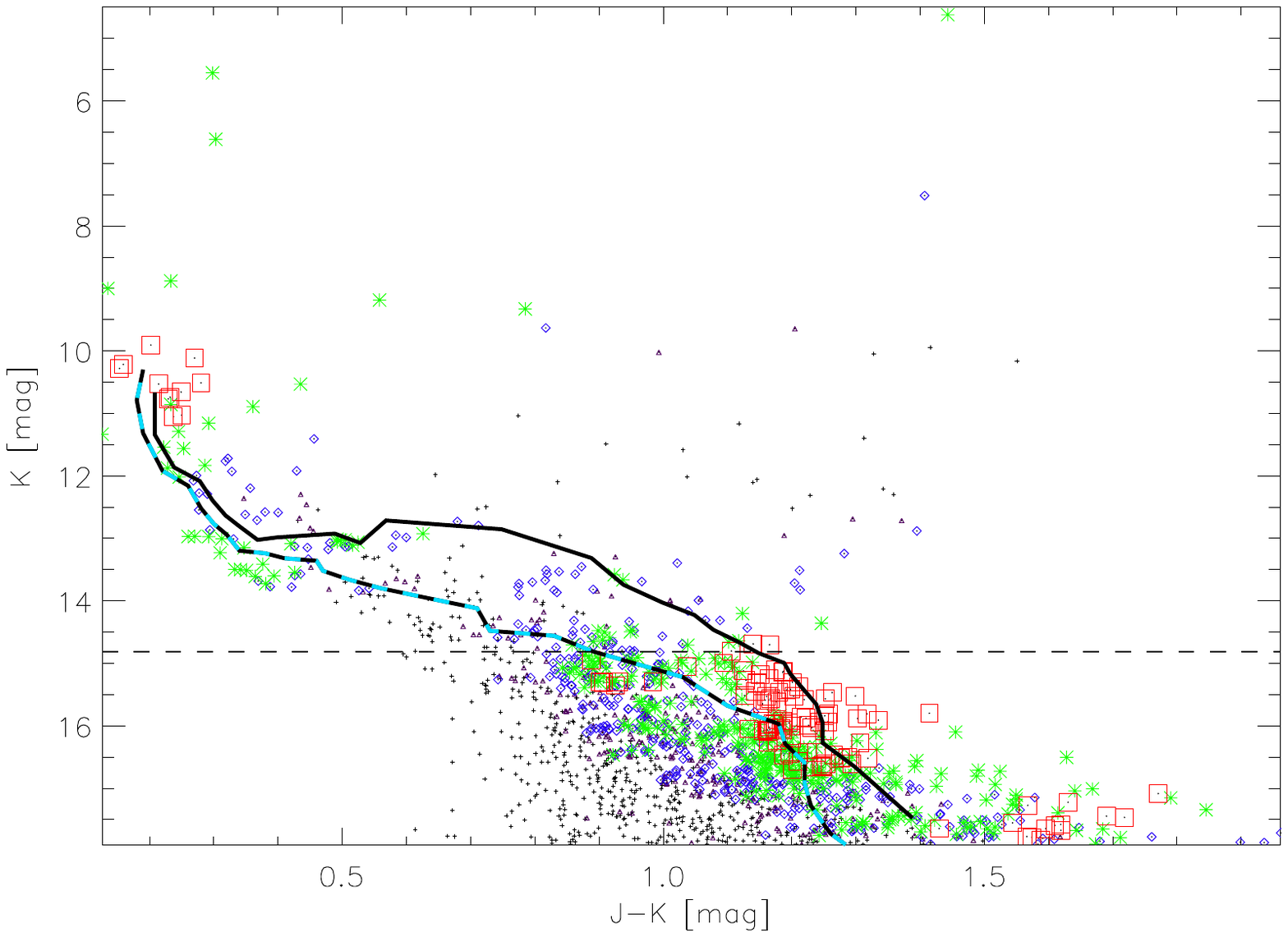} \hfill
        \includegraphics[width=\columnwidth,viewport=54 360 558 720]{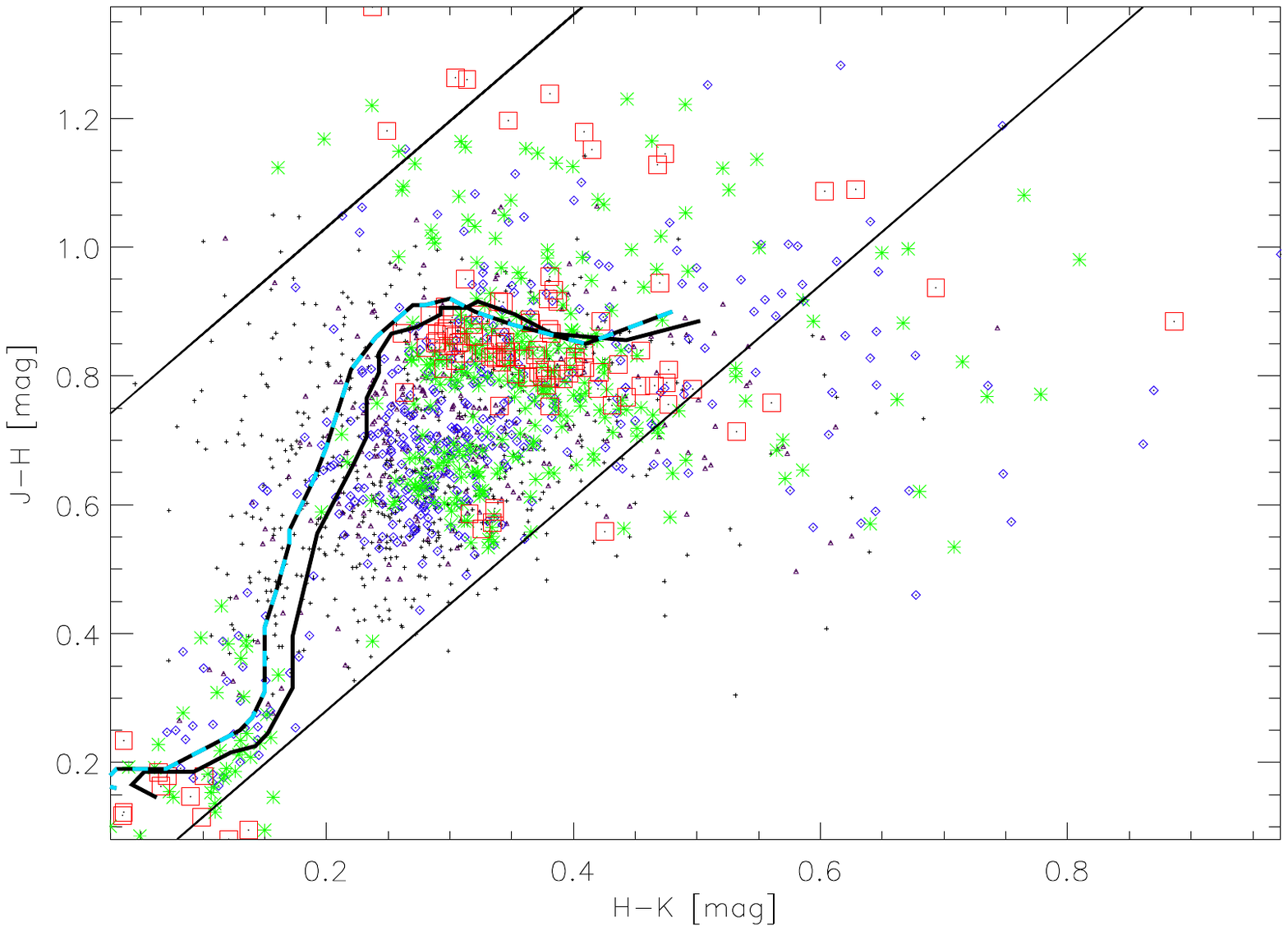}\\
        \caption{\label{fig_ccm_fsr0870} As in Figure\,\ref{fig_ccm_fsr0089}, but for \textbf{FSR0870}.}
\end{figure*}

\subsection{FSR0794} \label{cluster_0794}

A previously known and extensively studied cluster (NGC\,1960/M36), located in
the second Galactic Quadrant in the Auriga constellation. Figure
\ref{fig_ccm_fsr0794} shows a young open cluster with well defined MS and PMS
tracks. The cluster's properties are redetermined as $d=\,1.20$\,kpc, age of
10\,Myr and $A_{H}=\,0.35$\,mag. After testing different variations we found the
extinction law that best fits the cluster is $\chi=0.75$. These revised age and
distance values are in agreement with those derived in Paper II, but the
extinction value is significantly larger caused by a change in the extinction
law for the cluster. The brightest member of the cluster listed in SIMBAD is a
B2 type star (MPI$=0.95$), which is consistent with our derived age value for
FSR0794.

This cluster has been extensively studied in the literature and the majority of
determined properties values agree with the here revised values, albeit with a
higher extinction value due to our use of a different extinction law than the
majority of the literature. As such, FSR0794 is a good candidate to demonstrate
the reliability of our isochrone fitting procedure to accurately establish
cluster properties. 

\subsection{FSR0828} \label{cluster_0828}

A cluster candidate  of the FSR list, located in the second Galactic Quadrant.
In Paper II FSR0828 was flagged as having the greatest Galactocentric distance
in the sample at $R_{GC}=13.0$\,kpc. Figure \ref{fig_ccm_fsr0828} shows an old
cluster with a well defined MS (below the 2MASS detection limit), turn-off and
giants. The cluster's properties are redetermined as $d=\,3.20$\,kpc, age of
2\,Gyr and $A_{H}=\,0.22$\,mag. As such, FSR0828 no longer has the largest
Galactocentric distance in the sample. After testing different variations we
found the extinction law that best fits the cluster is $\chi=1.55$. These
revised values make the cluster nearer and older than the values derived in
Paper II suggest, but are in good agreement with the literature. 

\subsection{FSR0870} \label{cluster_0870}

A previously known cluster (NGC\,2129), located in the third Galactic Quadrant
in the Gemini constellation. Figure \ref{fig_ccm_fsr0870} clearly show a very
young open cluster with a PMS track. The cluster's properties are redetermined
as $d=\,1.45$\,kpc, age of 10\,Myr and $A_{H}=\,\,0.62$\,mag. After testing
different variations we found the extinction law that best fits the cluster is
$\chi=1.65$. These revised values make the cluster older, redder and nearer than
determined in Paper II. The brightest member of the cluster listed in SIMBAD is
a B2 type star (MPI$=0.75$), which is consistent with our derived age value for
FSR0870.

The nature of FSR0870 has been a subject for debate since its discovery. An
early study by \citet{1938AnHar.106...39C} placed the cluster at a distance of
$\sim$\,0.6\,kpc from the Sun, but a study by \citet{1994RMxAA..28..139P} cast
doubt on the cluster's existence when the authors created a histogram of the
distances of 37 stars in the direction of FSR0870 using $uvby-\beta$ photometry
and concluded it was an asterism, despite a previous radial velocity study by
\citet{1991AJ....102.1103L}. Their conclusion prompted further studies which
utilised proper motions and photometry confirm whether FSR0870 is real or an
asterism (e.g. \citet{2000A&AS..146..251B}, \citet{2006MNRAS.365..867C}). It is
now generally accepted that FSR0870 is a real Solar metallicity cluster, at a
distance of $\sim$\,2\,kpc, age of $\sim$\,10\,Myr and extinction of
$A_{H}\sim\,0.40$ mag. However, it should be noted that the majority of the MS
is below the 2MASS (and previous studies) detection limit.

\begin{figure*}
        \includegraphics[width=\columnwidth,viewport=54 360 558 720]{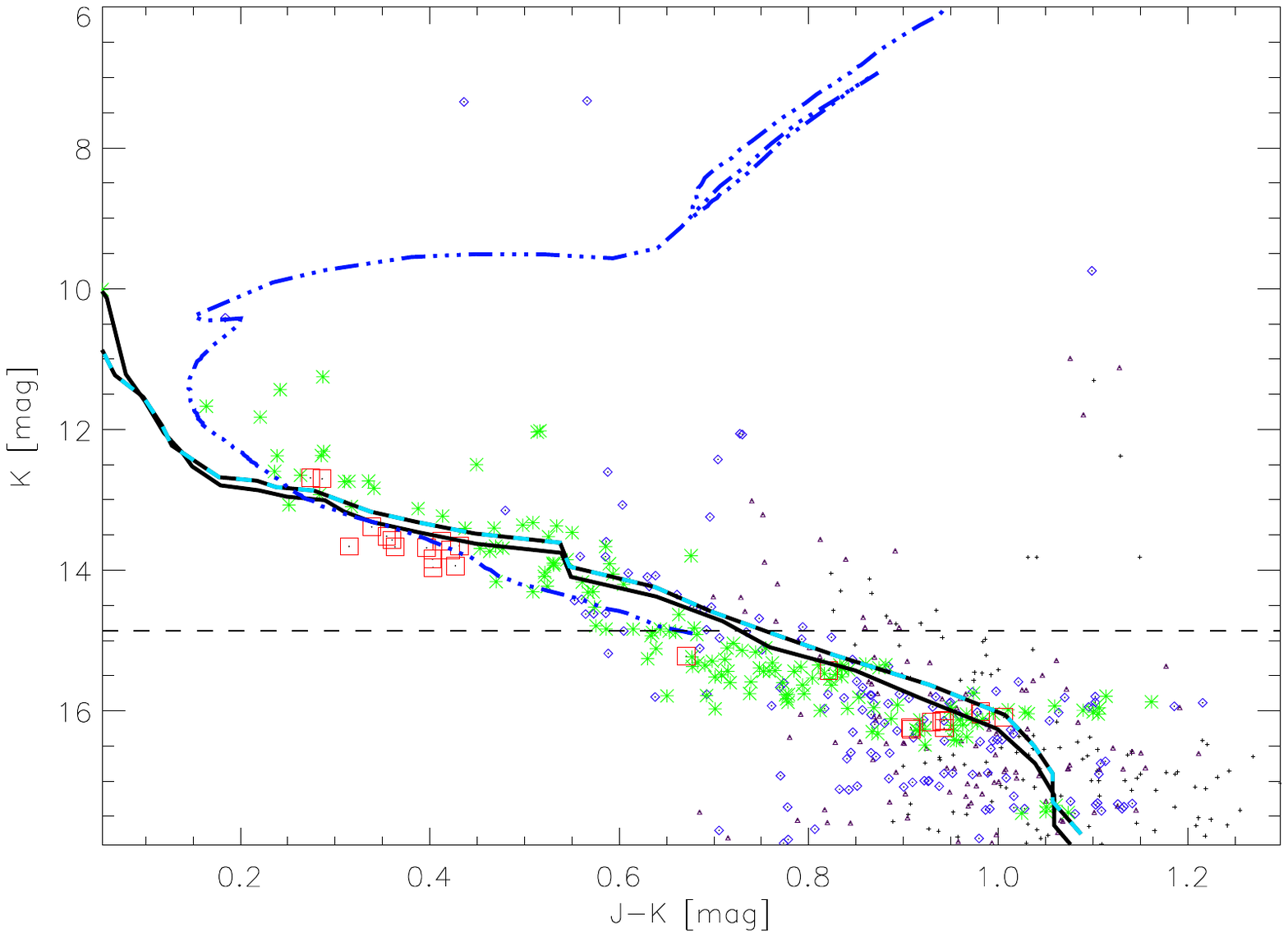}\hfill
        \includegraphics[width=\columnwidth,viewport=54 360 558 720]{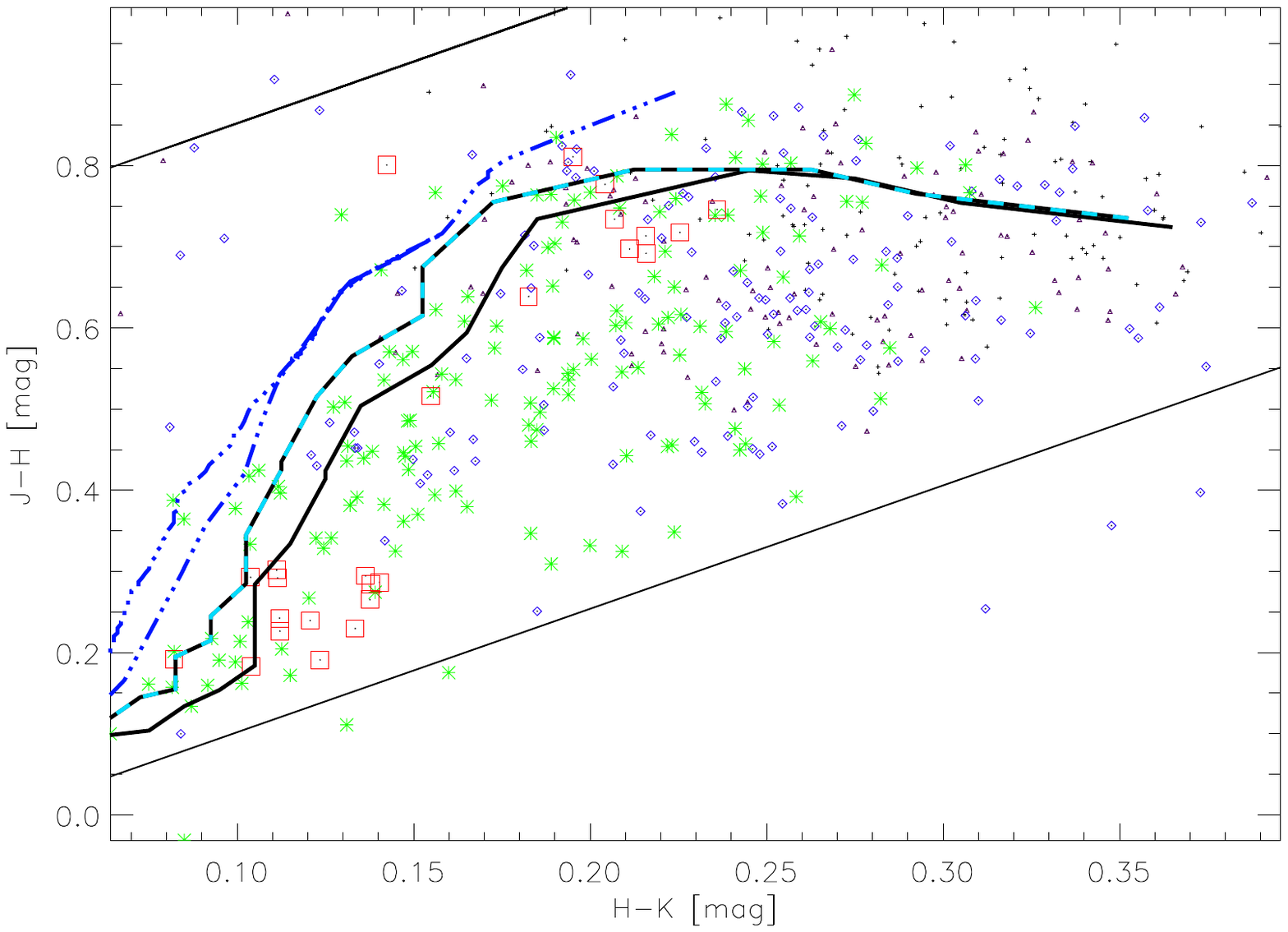}\\
        \caption{\label{fig_ccm_fsr0904} As in Figure\,\ref{fig_ccm_fsr0089}, but for \textbf{FSR0904}. The triple-dot-dash blue isochrone represents the best fit as determined by \citet{2010AstL...36...75G}.}
\end{figure*}

\begin{figure*}
        \includegraphics[width=\columnwidth,viewport=54 360 558 720]{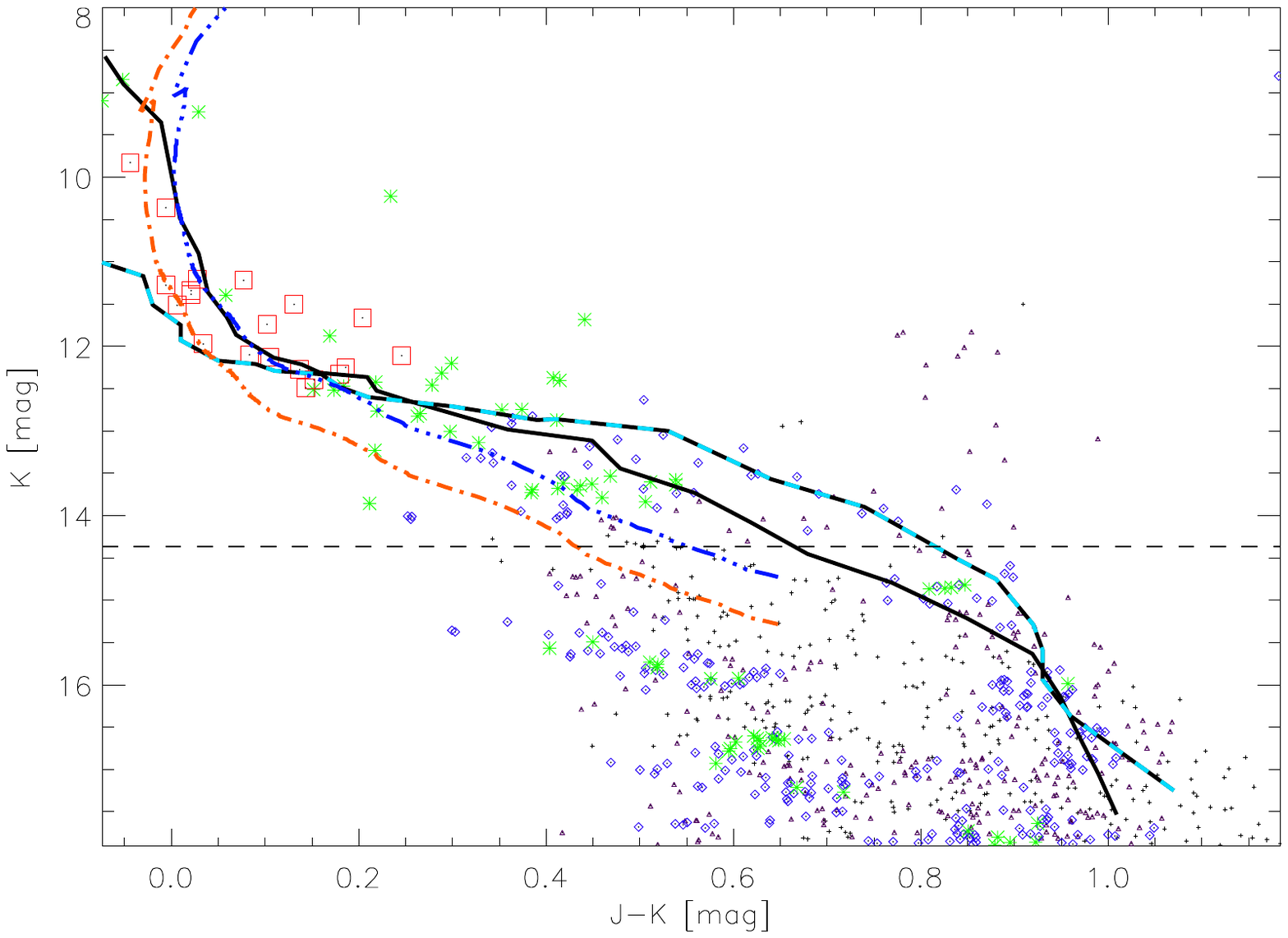}\hfill
        \includegraphics[width=\columnwidth,viewport=54 360 558 720]{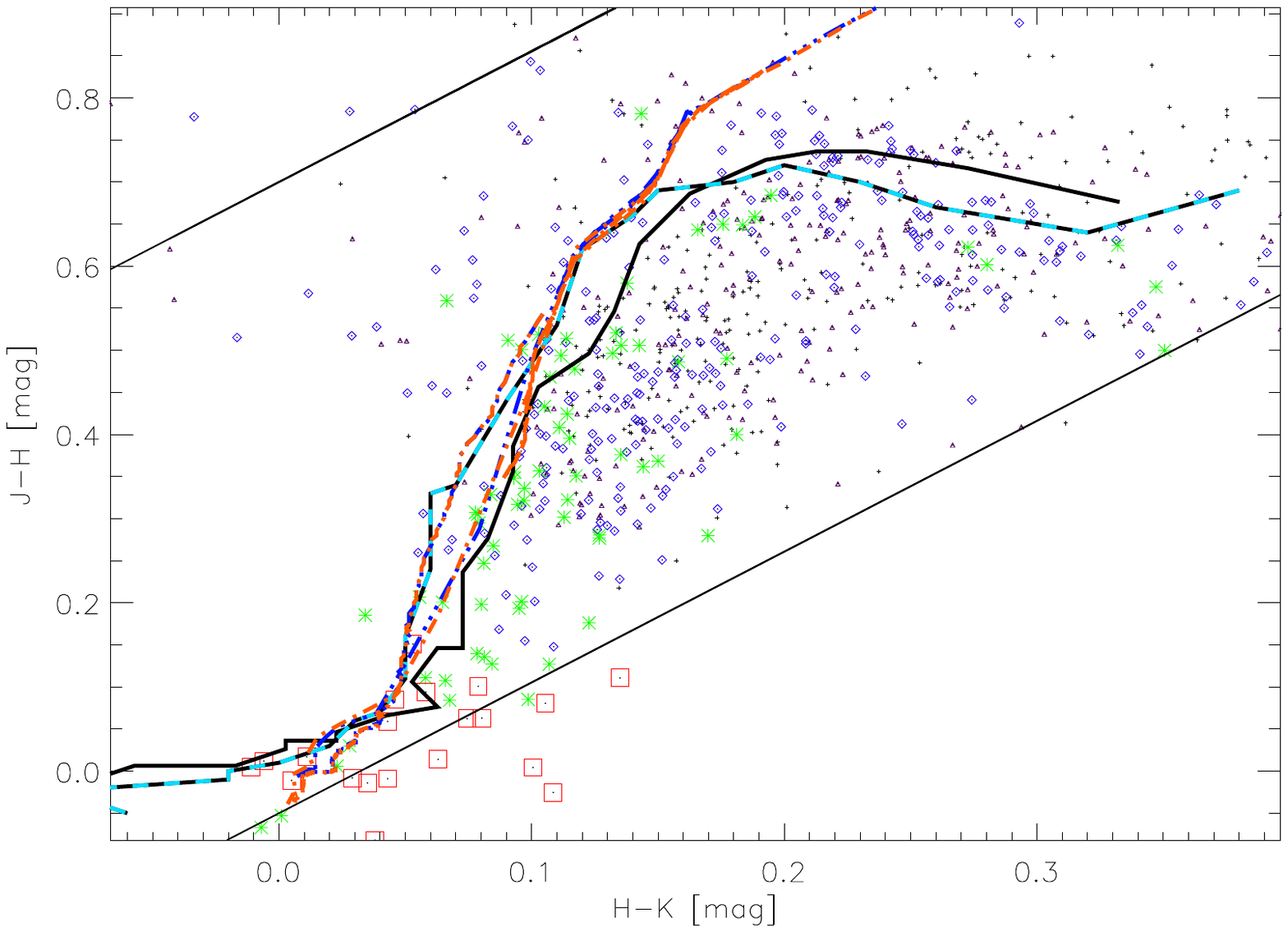}\\
        \caption{\label{fig_ccm_fsr1189} As in Figure\,\ref{fig_ccm_fsr0089}, but for \textbf{FSR1189}. The dot-dash orange and triple-dot-dash blue isochrones represent the best fit as determined by \citet{2014RMxAA..50...15S} and \citet{2011JKAS...44...39L} respectively.}
\end{figure*}

\subsection{FSR0904} \label{cluster_0904}

A cluster candidate located in the third Galactic Quadrant. Figure
\ref{fig_ccm_fsr0904} shows a young open cluster without the PMS track it was
suspected to have. The cluster's properties are redetermined as
$d=\,\,1.50$\,kpc, age of 80\,Myr and $A_{H}=\,0.30$\,mag. After testing
different variations we found the extinction law that best fits the cluster is
$\chi=1.52$. These revised values are in general agreement with the values
derived in Paper II, albeit with slightly higher age, lower extinction and
distance, estimates due to the majority of the MS falling below the 2MASS
detection limit. The brightest star in the cluster area (listed in SIMBAD) is a
B1 star, that would suggest that the cluster is much younger than our derived
value. However it has a relatively low MPI of $0.55$, so it is uncertain whether
this star is a true cluster member or a field star and a spectroscopic analysis
is required to determine whether it is a cluster member (which is outside the
scope of this paper).

There have been three previous studies of FSR0904, and whilst all have derived
values for the clusters properties by fitting isochrones to 2MASS photometry,
they have produced conflicting values. The revised values are in agreement with
those derived by \citet{2010A&A...521A..42C}, but conflict with those of
\citet{2010AstL...36...75G} and the MWSC catalogue.

\citet{2010AstL...36...75G} found the cluster to be a factor of $\sim$\,10 older
than the revised values suggest, with a significantly smaller extinction and
distance. This discrepancy is caused by misidentification of cluster members by
the authors (i.e. field stars outside $2\,\times\,r_{cor}$ as giants - see
Figure\,\ref{fig_ccm_fsr0904}), which was compounded by the majority of the MS
not being visible to the authors when they determined the cluster's properties
(as it is below the 2MASS detection limit). Members were identified as objects
that lay along the isochrone fitted to the CCM diagrams and which formed the
cluster's spatial density peak. Obviously, this approach is subjective to the
authors interpretation of the CCM diagrams and subsequent choice of isochrone.
Furthermore, assessing membership based on spatial positioning has been shown to
be unreliable, especially for young clusters (such as FSR0904) which do not
appear circular in projection, those projected onto a high density field star
background, or with significant stellar crowding (for a full discussion see e.g.
\citet{2010MNRAS.409.1281F}, \citet{2013MNRAS.436.1465B}). Although
\citet{2010AstL...36...75G} identified the majority of members on the top of the
MS with reasonable accuracy, the misidentification of field stars as the
cluster's giants, in combination with `missing' the majority of the cluster
sequence (due to 2MASS detection limits), has resulted in the authors
significantly overestimating FSR0904's age and underestimating its extinction
and distance.


\subsection{FSR1189} \label{cluster_1189}

A previously known cluster (NGC\,2353), located in the third Galactic Quadrant
near the Canis Major OB1 association. Figure \ref{fig_ccm_fsr1189} shows a
moderately young open cluster with a PMS track. The cluster's properties are
redetermined as $d=\,1.20$\,kpc, age of 90\,Myr and $A_{H}=\,0.11$\,mag. After
testing different variations we found the extinction law that best fits the
cluster is $\chi=1.55$. These revised values are in agreement with the values
derived in Paper II. 

The revised age and distance values are also in consensus with the literature.
This age also agrees with the assessment of \citet{1990PASP..102..865F} and
\citet{2011JKAS...44...39L}, that FSR1189 is unrelated to the much younger
($\sim$\,3\,Myr), but similarly distanced, Canis Major OB1 association. The
revised extinction estimate is a factor of 2 larger than that given by various
authors, but is accurate according to the isochrone fits shown in
Figure\,\ref{fig_ccm_fsr1189}. Literature studies have predominantly analysed
FSR1189 using visual UBV photometry (see e.g. \citet{2001A&A...370..436M}),
whilst this work has conducted the analysis using NIR JHK photometry. 

\begin{figure*}
        \includegraphics[width=\columnwidth,viewport=54 360 558 720]{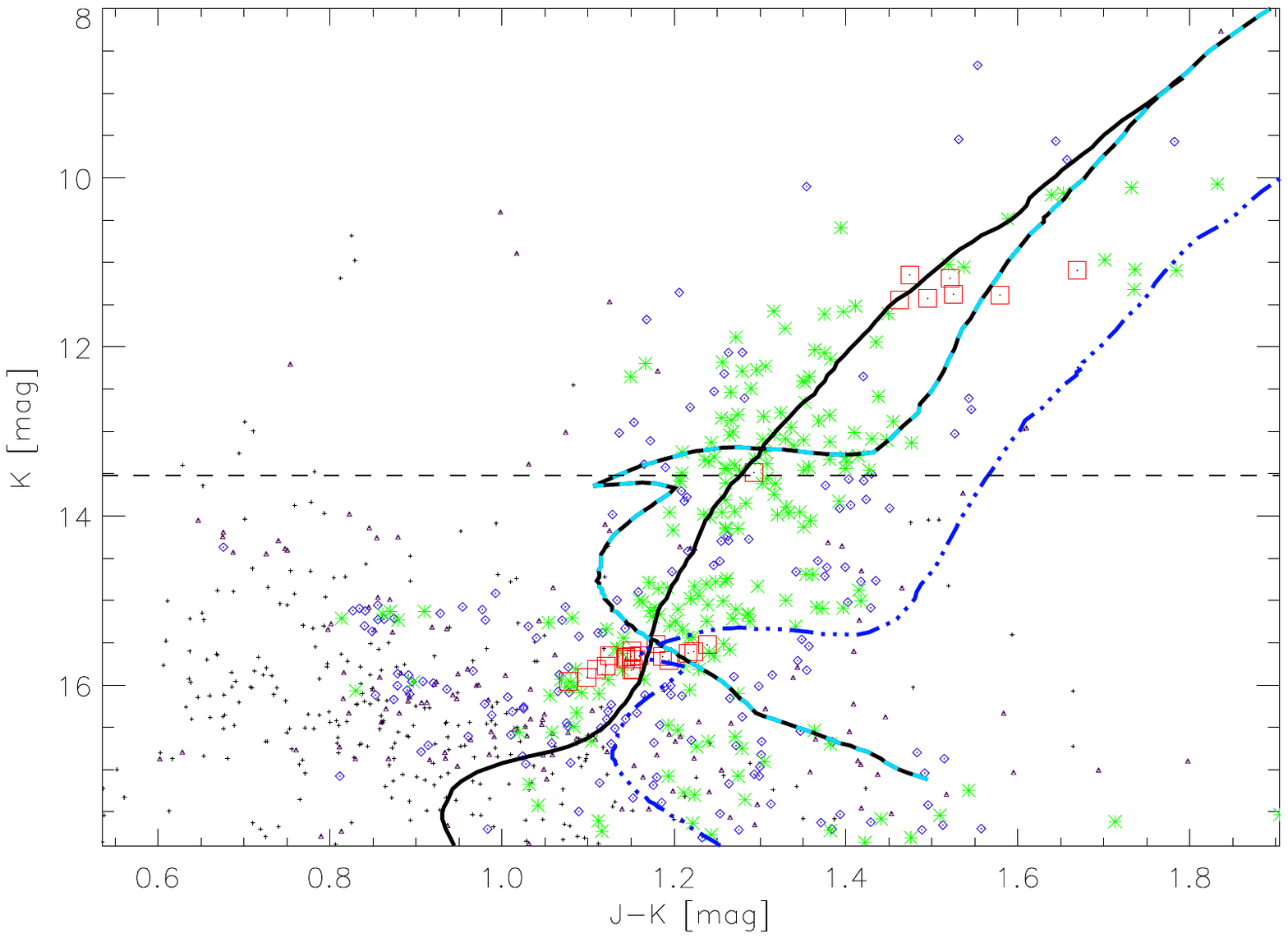}\hfill
        \includegraphics[width=\columnwidth,viewport=54 360 558 720]{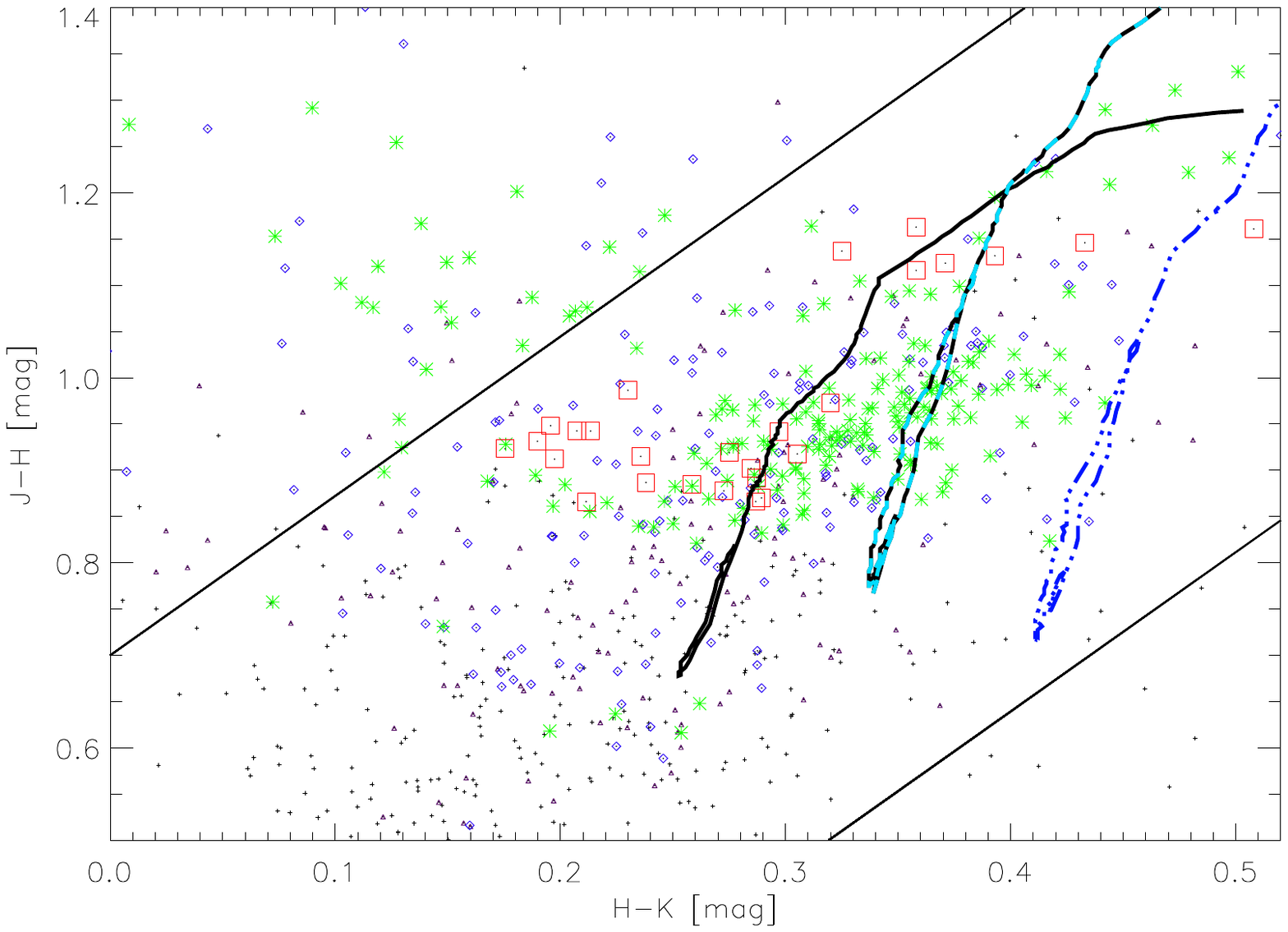}\\
        \caption{\label{fig_ccm_fsr1716} As in Figure\,\ref{fig_ccm_fsr0089}, but for \textbf{FSR1716}. The triple-dot-dash blue isochrones represent the best fitting open cluster isochrone as determined \citet{2010MNRAS.409.1281F}.}
\end{figure*}
\subsection{FSR1716} \label{cluster_1716}

A confirmed cluster candidate of the FSR list, located in the fourth Galactic
Quadrant. In Paper II FSR1716 was flagged as one of the nearest clusters to the
GC in the FSR list at $R_{GC}=4.3$\,kpc. Figure \ref{fig_ccm_fsr1716} shows a
very old open or possible globular cluster candidate. The cluster's properties
are redetermined as $d=\,7.30$\,kpc, age of 10-12\,Gyr and $A_{H}=\,0.57$\,mag.
After testing different variations we found the extinction law that best fits
the cluster is $\chi=1.72$. These revised values make the cluster older, further
away but less reddened than the values in Paper II suggest. The revised values
make FSR1716 both the oldest cluster in the sample and nearest to the GC with
$R_{GC}=4.1$\,kpc. 

Deriving the age of old clusters is difficult. Unlike younger clusters (for
which the isochrone can be fitted to the MS stars, turn-off and potentially a
few giants) the CCM diagrams of the oldest clusters have only one detectable age
defining feature: the position of the Red Giant ``Clump" (RGC) i.e. horizontal
branch stars expected to have degenerated into a clump near the Red Giant
Branch. Unfortunately in many instances this feature falls below, or is very
close to, the detection limit of the photometry being employed to analyse the
cluster, and its position has to be guesstimated by each author based on their
individual interpretations of the CCM diagrams. If FSR1716 had an age $<$10\,Gyr
we would expect to detect the MS turn-off stars but as
Figure\,\ref{fig_ccm_fsr1716} clearly shows this is not the case. Instead there
appears to be a prominent RGC at $K\,\sim\,15$\,-\,$15.5$\,mag, evidencing its
status as an older cluster.

\begin{figure}
        \includegraphics[width=\columnwidth,viewport=54 360 558 720]{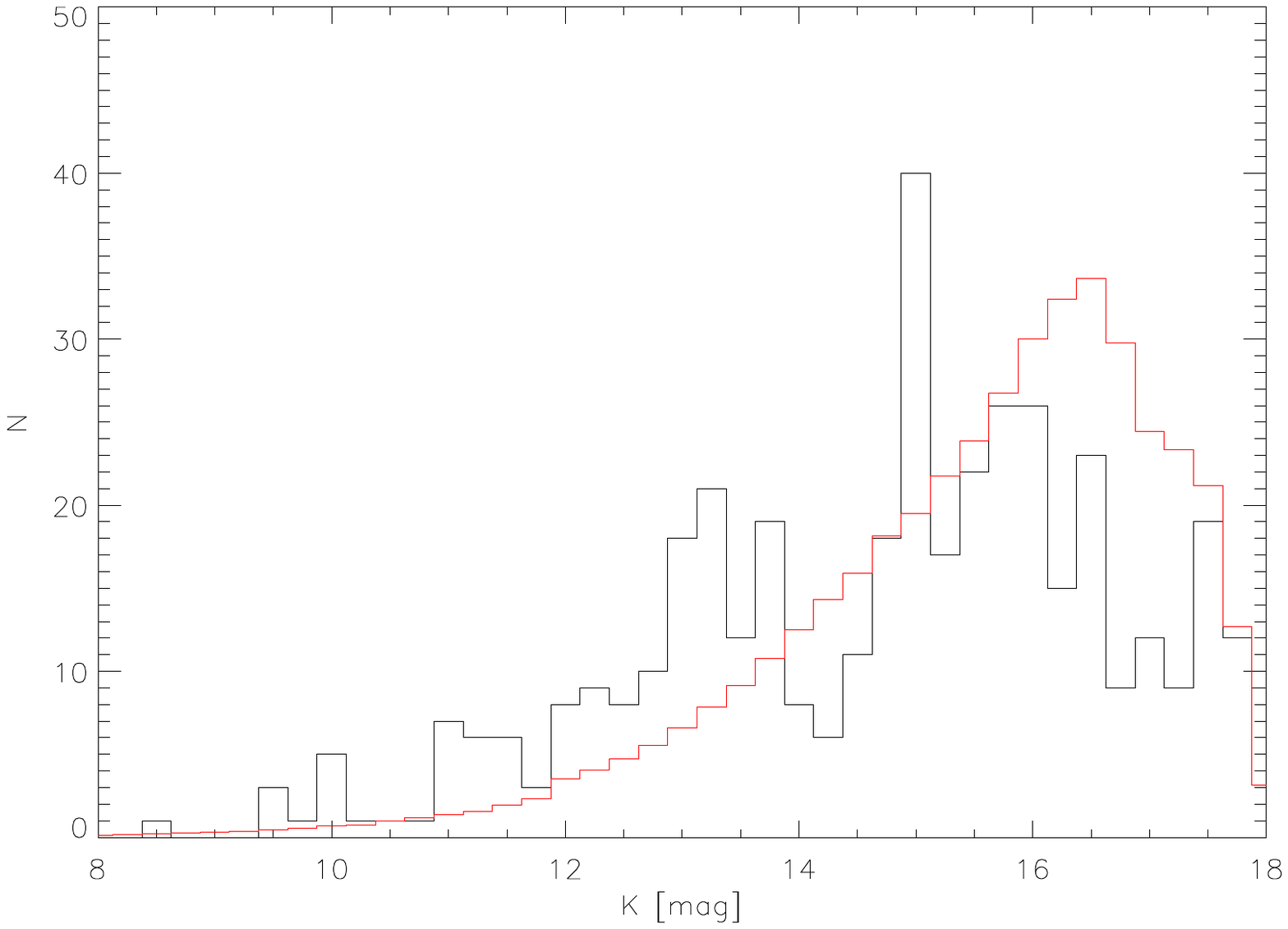}\hfill
        \caption{\label{fig_histkmag_fsr1716} Histograms of the $K$-band magnitudes of \textbf{FSR1716} for stars (i) within $2\,\times\,r_{cor}$ and $P^i_{cl}\ge\,40\%$ ($N=15$), represented by the solid black line; and (ii) the control field (normalised to the cluster area), represented by the solid red line. The peak in the cluster members histogram at $K\sim15$\,mag indicates the position of the cluster's RGC.}
\end{figure}

To derive an age estimate for FSR1716 we determine the exact position of the
RGC, derived as the weighted mean of the membership probabilities of stars in
the cluster area, determined to be $K\,=$\,15.3\,mag. To confirm, we plot a
histogram of the $K$-band magnitude of stars with a MPI of $>\,0.4$ in the
cluster area and all stars in the control field (normalised to the cluster
area). Figure \ref{fig_histkmag_fsr1716} clearly shows that there is a
significant peak in the histogram of cluster members at $K\,\sim\,15$\,mag
(which is independent of bin size) above that of the histogram of control field
stars. Furthermore, there is a $\sim\,5.5$ magnitude difference between the top
of the RGB and the peak. Thus we interpret the peak at $K\,\sim\,15.3$\,mag as
the RGC. Finally, we find a second significant peak at $K\,\sim\,13$\,mag
(independent of bin size) which we interpret as the Asymptotic Giant ``Clump"
i.e. Asymptotic Giant Branch stars at the onset of helium shell burning
(\citet{2002ApJ...573L..51A}, \citet{2007MNRAS.377L..54F}).

The position of the RGC is fully/partially below the detection limits of the
photometry which has been previously used to derive the cluster's fundamental
properties (2MASS, NTT). As such there has been no agreement in the literature
of the age and distance of FSR1716, further complicated by a debate over its
nature. Most studies have assumed that FSR1716 is an open cluster, deriving ages
of $1.7$\,-\,$7.1$\,Gyr and distances of $0.8$\,-\,$7.0$\,kpc. Meanwhile,
\citet{2008A&A...491..767B} concluded that it could be a low metallicity
$12$\,Gyr globular cluster at a distance of $2.3$\,kpc. 

The revised property values presented here were determined by fitting an
isochrone to the CCM diagrams of FSR1716, using the measured position of the RGC
as a reference point. Our age value of FSR1716 agrees with the assessment that
it is a potential globular cluster candidate.

\section{Conclusions}\label{sect_conclude}

We selected thirteen FSR list clusters from \citet{2014MNRAS.444..290B} and
analysed their properties using deep, high resolution photometry from the
UKIDSS-GPS and VISTA-VVV surveys. These clusters were selected for our sample as
they were either (i) suspected to either contain a significant number of PMS
stars or (ii) had a previously determined notable Galactocentric distance. Of
these, seven were confirmed to contain PMS stars, one of which is a confirmed
new cluster candidate. Notably, the analysis identified FSR1716 as a globular
cluster candidate with a distance of about 7.3\,kpc and an age of
10\,--\,12\,Gyr.

For the majority of our selected clusters this was the first analyses of them
which used deep, high resolution photometry, and as such their derived
properties differed substantially from literature estimates. For the majority of
these clusters there was a marked discrepancy between the properties derived in
this study (and the literature) with those listed in the MWSC catalogue, which
is most likely caused by the nature of these clusters the pipeline employed by
the authors of the MWSC catalogue to homogeneously derive cluster properties
using 2MASS photometry, as they assumed a constant extinction law ($\chi=2.00$;
\citet{2012A&A...543A.156K}). This, compounded by the absence of deep, high
resolution photometry, has resulted in some erroneous values for individual
clusters. 

It is important to determine the properties of the remaining FSR list clusters
and confirm the nature of the new cluster candidates which could not be studied
in this series of papers due to a lack of available deep, high resolution
photometry. Unfortunately, to date surveys such as those used in this study
(UKIDSS-GPS and VISTA-VVV) are only complete for a fraction of FSR list
clusters. Mass estimates for the clusters of the catalogue would enrich the
understanding of observed cluster distributions in the Galactic Plane.

\section*{Acknowledgements}\label{sect_ack}

A.S.M. Buckner acknowledges a Science and Technology Facilities Council
studentship and a University of Kent scholarship. This publication makes use of
data products from the Two Micron All Sky Survey, which is a joint project of
the University of Massachusetts and the Infrared Processing and Analysis
Center/California Institute of Technology, funded by the National Aeronautics
and Space Administration and the National Science Foundation.  This work is
based in part on data obtained as part of the UKIRT Infrared Deep Sky Survey.
The VVV Survey is supported by the European Southern Observatory,  by BASAL
Center for Astrophysics and Associated Technologies PFB-06,  by FONDAP Center
for Astrophysics 15010003, by the Chilean Ministry  for the Economy,
Development, and Tourism's Programa Iniciativa  Cient\'{\i}fica Milenio through
grant P07-021-F, awarded to The  Milky Way Millennium Nucleus. This publication
makes use of data products from the Wide-field Infrared Survey Explorer, which
is a joint project of the University of California, Los Angeles, and the Jet
Propulsion Laboratory/California Institute of Technol-ogy,  funded by the
National Aeronautics and Space Administration. This  research has made use of
the WEBDA database, operated  at the Institute for Astronomy of the University
of Vienna. This research has made use of the SIMBAD database, operated at CDS,
Strasbourg, France.



\bibliographystyle{mn2e}
\bibliography{refs} 
\clearpage

\onecolumn
\appendix
\section{A table of clusters}
\begin{longtable}{|l|l|l|l|l|l|l|l|l|}
\caption{\label{obsinterest_results} Details the refined properties of the clusters studied. For each cluster the table lists its ID, class (PMS, OC or GlC), type (known open cluster or new cluster candidate), Galactic coordinates (l,b), age, distance in parsec, $H$-band extinction value, and the literature source of these values.} \\
\hline 

\multicolumn{1}{|c|}{\textbf{ID}} &
\multicolumn{1}{c|}{\textbf{Type}} &
\multicolumn{1}{c|}{\textbf{Class}} &
\multicolumn{1}{c|}{\textbf{l}} &
\multicolumn{1}{c|}{\textbf{b}} &
\multicolumn{1}{c|}{\textbf{Age}} &
\multicolumn{1}{c|}{\textbf{d}} &
\multicolumn{1}{c|}{\textbf{$A_H$}} &
\multicolumn{1}{c|}{\textbf{Reference}}\\
\multicolumn{1}{|c|}{} &
\multicolumn{1}{c|}{} &
\multicolumn{1}{c|}{} &
\multicolumn{1}{c|}{[deg]} &
\multicolumn{1}{c|}{[deg]} &
\multicolumn{1}{c|}{[log(age/yr)]} &
\multicolumn{1}{c|}{[pc]} &
\multicolumn{1}{c|}{[mag]} &
\multicolumn{1}{c|}{}\\
\hline  \endfirsthead

\multicolumn{9}{c}%

 {{\bfseries \tablename\ \thetable{} -- continued from previous page}}     \\ \hline

\multicolumn{1}{|c|}{\textbf{ID}} &
\multicolumn{1}{c|}{\textbf{Type}} &
\multicolumn{1}{c|}{\textbf{Class}} &
\multicolumn{1}{c|}{\textbf{l}} &
\multicolumn{1}{c|}{\textbf{b}} &
\multicolumn{1}{c|}{\textbf{Age}} &
\multicolumn{1}{c|}{\textbf{d}} &
\multicolumn{1}{c|}{\textbf{$A_H$}} &
\multicolumn{1}{c|}{\textbf{Reference}}\\
\multicolumn{1}{|c|}{} &
\multicolumn{1}{c|}{} &
\multicolumn{1}{c|}{} &
\multicolumn{1}{c|}{[deg]} &
\multicolumn{1}{c|}{[deg]} &
\multicolumn{1}{c|}{[log(age/yr)]} &
\multicolumn{1}{c|}{[pc]} &
\multicolumn{1}{c|}{[mag]} &
\multicolumn{1}{c|}{}\\
\hline  \endhead

\multicolumn{9}{|r|}{{Continued on next page}}     \\ \hline
\endfoot

\endlastfoot
FSR0089 &New&OC&  29.49  & -00.98 & 8.70 & 3100 & 1.53 &$^{1}$\\
 & & & & & 8.50 & 6500$^{+70}_{-70}$ & 1.50 &$^{2}$\\
 & & & & & 9.00 & 2200$^{+100}_{-100}$ & 1.60 $^{+0.05}_{-0.05}$ & $^{3}$\\
 & & & & & 9.00 & 2200$^{+300}_{-300}$ & 1.63 $^{+0.08}_{-0.08}$ & $^{4}$\\
MWSC2997 & & & & & 8.30 & 1516 & 1.19 & $^{5}$\\
\hline
FSR0188 &New&OC&  70.65  &  01.74 & 9.00 & 4900 & 0.61 &$^{1}$\\
 & & & & & 8.60 & 10500$^{+1000}_{-1000}$ & 0.62$^{+0.02}_{-0.02}$ &$^{2}$\\
MWSC3228 & & & & & 8.20 & 3354 & 0.84 &$^{5}$\\
\hline
FSR0195 &New&PMS&  72.07  & -00.99 & 7.48 & 3500 & 1.25 &$^{1}$\\
 & & & & & 7.60 & 1900 & 1.15 &$^{2}$\\
MWSC3298 & & & & & 9.35 & 2331 & 0.92 &$^{5}$\\
\hline
FSR0207 &Known&PMS&  75.38  &  01.30 & 7.00 & 1400 & 0.25 &$^{1}$\\
 & & & & & 7.00 & 1400 & 0.30 &$^{2}$\\
MWSC3297 & & & & & 7.15 & 1764 & 0.33 & $^{5}$\\
IC\,4996 & & & & & 6.95 & 1620$^{+75}_{-75}$ & 0.34 & $^{6}$\\
 & & & & & 6.88 & 2399 & 0.39$^{+0.04}_{-0.04}$ & $^{7}$\\
 & & & & & 6.85 & 2291$^{+995}_{-995}$ & 0.38 & $^{8}$\\
 & & & & & 7.00 & 1620 & 0.35 & $^{9}$\\
\hline
FSR0301 &Known&PMS&  93.04  &  01.80 & 7.70 & 2250 & 0.87 & $^{1}$\\
 & & & & & 7.51 & 2000 & 1.00 &$^{2}$\\
MWSC3490 & & & & & 8.57 & 1700 & 1.02 & $^{5}$\\
Berkeley\,55 & & & & & 7.70 & 4000 & 1.00$^{+0.09}_{-0.09}$ & $^{10}$\\
 & & & & & 8.48 & 1440$^{+65}_{-65}$ & 0.81 & $^{11}$\\
 & & & & & 8.50 & $1210^{+310}_{-390}$ & $ 0.94^{+0.05}_{-0.06}$ & $^{12}$\\
\hline
FSR0636 &Known&PMS&  143.35 & -00.13 & 7.78 & 720 & 0.33 &$^{1}$\\
& & & & & 7.70 & 800 & 0.35 &$^{2}$\\
MWSC0277 & & & & & 8.98 & 632 & 0.19 &$^{5}$\\
King\,6 & & & & & 8.40 & 871$^{+12}_{-12}$ & 0.27$^{+0.05}_{-0.05}$ &$^{13}$\\
 & & & & & 8.40 & $800^{+290}_{-250}$ & $ 0.29^{+0.07}_{-0.06}$ &$^{12}$\\
\hline
FSR0718 &Known&OC&  162.27 &  01.62 & 7.90 & 2400 & 0.45 &$^{1}$\\
& & & & & 7.30 & 2700 & 0.55 &$^{2}$\\
MWSC0453 & & & & & 9.40 & 1300 & 0.22 &$^{5}$\\
Berkeley\,15 & & & & & 8.50$^{+0.10}_{-0.10}$ & 3000$^{+300}_{-300}$ & 0.48$^{+0.03}_{-0.03}$ &$^{14}$\\
 & & & & & 9.70 & 1259$^{+135}_{-135}$ & 0.25 &$^{15}$\\
 & & & & & 9.35/9.95$^{+0.05}_{-0.05}$ & 1406$^{+10}_{-10}$ & 0.13$^{+0.02}_{-0.02}$ &$^{16}$\\
\hline
FSR0794 &Known&PMS&  174.55 &  01.08 & 7.00 & 1200 & 0.35&$^{1}$\\
 & & & & & 7.30 & 1200 & 0.20 &$^{2}$\\
MWSC0594 & & & & & 7.57 & 1200 & 0.16 & $^{5}$\\
NGC\,1960 & & & & & 7.40 & 1330 & 0.12 & $^{17}$\\
 & & & & & 7.20 & 1318$^{+120}_{-120}$ & 0.14$^{+0.01}_{-0.01}$ & $^{18}$\\
 & & & & & 7.48 & 1200 & 0.13 & $^{19}$\\
M\,36 & & & & & 7.42 & $1164^{+11}_{-26}$ & 0.11 & $^{20}$\\
\hline
FSR0828 &New&OC&  179.92 &  01.75 & 9.30 & 3200 & 0.22 &$^{1}$\\
 & & & & & 8.90 & 5000 & 0.28 &$^{2}$\\
MWSC0687 & & & & & 9.12 & 3000 & 0.42 & $^{5}$\\
 & & & & & 9.30 & 2800 & 0.28 & $^{4}$\\
Koposov\,43 & & & & & 9.30 & 2800$^{+120}_{-120}$ & 0.21$^{+0.05}_{-0.05}$ & $^{21}$\\
\hline
\clearpage
FSR0870 &Known&PMS&  186.61 &  00.15 & 7.00 & 1450 & 0.62 &$^{1}$\\
 & & & & & 7.30 & 1600 & 0.40 &$^{2}$\\
MWSC0704 & & & & & 7.48 & 1651 & 0.48 & $^{5}$\\
NGC\,2129 & & & & & 7.00 & 2200$^{+200}_{-200}$ & 0.43$^{+0.04}_{-0.04}$ & $^{22}$\\
 & & & & & 7.00 & 2100$^{+100}_{-100}$ & 0.42$^{+0.03}_{-0.03}$ & $^{23}$\\
 & & & & & 7.10 & 1950 & 0.45 & $^{24}$\\
\hline
FSR0904 &New&OC&  191.03 & -00.78 & 7.90 & 1500 & 0.30 & $^{1}$\\
 & & & & & 7.30 & 2000 & 0.43 &$^{2}$\\
MWSC0731 & & & & & 7.80 & 1427 & 0.25 & $^{5}$\\
SAI61 & & & & & 7.30 & 2200$^{+100}_{-100}$ & 0.35$^{+0.02}_{-0.02}$ & $^{25}$\\
 & & & & & 8.80 & 1265$^{+10}_{-10}$ & 0.08$^{+0.09}_{-0.09}$ & $^{26}$\\
\hline
FSR1189 &Known&PMS&  224.67 &  00.40 & 7.95 & 1200 & 0.11 & $^{1}$\\
 & & & & & 8.00 & 1200 & 0.10 &$^{2}$\\
MWSC1152 & & & & & 7.20 & 1180 & 0.08 & $^{5}$\\
NGC\,2353 & & & & & 8.10 & 1170$^{+40}_{-40}$ & 0.05$^{+0.01}_{-0.01}$ & $^{27}$\\
 & & & & & 7.88 & 1200$^{+80}_{-80}$ & 0.06 & $^{28}$\\
 & & & & & 7.86 & 1513$^{+646}_{-646}$ & 0.05$^{+0.03}_{-0.03}$ & $^{29}$\\
\hline
FSR1716 &New&OC/GlC?&  329.79 & -01.59 & 10.00-10.10 & 7300 & 0.57 & $^{1}$\\
 & & & & & 9.10 & 5400 & 0.79$^{+0.03}_{-0.03}$ &$^{2}$\\
 & & & & & $>9.30$ & 7000$^{+500}_{-500}$ & 0.93$^{+0.08}_{-0.08}$ &$^{4}$\\
 & & & & & 9.30 & 7000 & 0.75 &$^{30}$\\
 & &OC& & & 9.85 & 800$^{+100}_{-100}$ & 1.11$^{+0.04}_{-0.04}$ &$^{31}$\\
 & &GlC& & & 10.08 & 2300$^{+300}_{-300}$ & 1.11$^{+0.07}_{-0.07}$ &$^{31}$\\
MWSC2359 & & & & & 9.23 & 2396 & 1.20 &$^{5}$\\
\hline
\end{longtable}

\section*{Table Notes}

$^{1}$ This Paper ; $^{2}$ \citet{2014MNRAS.444..290B}; $^{3}$
\citet{2007A&A...473..445B}; $^{4}$ \citet{2008MNRAS.390.1598F}; $^{5}$
\citet{2013A&A...558A..53K}; $^{6}$ \citet{1996BaltA...5..539V}; $^{7}$
\citet{1998AJ....116.1801D}; $^{8}$ \citet{2007BASI...35..383B}; $^{9}$
\citet{1995yCat.7092....0L}; $^{10}$ \citet{2012AJ....143...46N}; $^{11}$
\citet{2008MNRAS.389..285T}; $^{12}$ \citet{2007A&A...467.1065M}; $^{13}$
\citet{2002AJ....123..905A}; $^{14}$ \citet{2004BASI...32..371L}; $^{15}$
\citet{2004BASI...32..295S}; $^{16}$ \citet{2010MNRAS.401..621T}; $^{17}$
\citet{2006AJ....132.1669S}; $^{18}$ \citet{2000A&A...357..471S}; $^{19}$
\citet{1985AZh....62..854B}; $^{20}$ \citet{2013MNRAS.434..806B}; $^{21}$
\citet{2008A&A...486..771K}; $^{22}$ \citet{2006MNRAS.365..867C}; $^{23}$
\citet{2013BASI...41..209T}; $^{24}$ \citet{2008A&A...486..771K}, paper
misidentified cluster as FSR\,0848; $^{25}$ \citet{2010A&A...521A..42C}; $^{26}$
\citet{2010AstL...36...75G}; $^{27}$ \citet{2011JKAS...44...39L}; $^{28}$
\citet{1990PASP..102..865F} ; $^{29}$ \citet{2014RMxAA..50...15S} ; $^{30}$
\citet{2010MNRAS.409.1281F} ; $^{31}$ \citet{2008A&A...491..767B} \\

\bsp	
\label{lastpage}
\end{document}